\documentclass[12pt]{article}
\usepackage{natbib}

\usepackage[english]{babel}
\usepackage{graphicx, color}
\usepackage{amsmath}
\usepackage{amssymb}
\usepackage{bm}

\begin{document}

\newcommand{\dd}{\mathrm{d}}
\newcommand{\rhob}{\rho_\text{b}}
\newcommand{\Tb}{T_\text{b}}
\newcommand{\kB}{\mbox{$k_\text{B}$}}
\newcommand{\rate}{\mbox{erg cm$^{-3}$ s$^{-1}$}}
\newcommand{\gcc}{\mbox{g~cm$^{-3}$}}
\newcommand{\xr}{x_\text{r}}
\newcommand{\gs}{g_\text{s}}
\newcommand{\me}{m_\text{e}}
\newcommand{\msun}{\mbox{$M_\odot$}}
\newcommand{\dr}[1]{\textcolor{red}{ #1}}
\newcommand{\dg}[1]{\textcolor{green}{ #1}}

\def\la{\;\raise0.3ex\hbox{$<$\kern-0.75em\raise-1.1ex\hbox{$\sim$}}\;}
\def\ga{\;\raise0.3ex\hbox{$>$\kern-0.75em\raise-1.1ex\hbox{$\sim$}}\;}

%

%
\begin{center}
\large{\bf {Traces of Anisotropic Quasi-Regular Structure in the SDSS Data}}
\end{center}

\begin{center}
{A.~I.~Ryabinkov} 
 and  {A.~D.~Kaminker}
\end{center}
\begin{center}
{Ioffe Physical Technical  Institute,  Politekhnicheskaya 26, 194021 St.~Petersburg, Russia} 
\end{center}

\begin{center}
{\it {kam.astro@mail.ioffe.ru}}\quad \quad {\it {Ryab@astro.ioffe.ru}}
\end{center}


 
%
\abstract{The aim of  this   study  is   
to search  for  quasi-periodical  structures at   
moderate cosmological  redshifts $z  \la 0.5 $. 
We  mainly use  the  SDSS DR7 data  
on   the  luminous red galaxies  (LRGs)
with  redshifts  $0.16 \leq z  \leq  0.47$.
At first,  we analyze 
features  (peaks) in the power spectra of radial (shell-like)
distributions using  separate angular  sectors  
in the sky  and calculate the power spectra within each sector.
As a result,  we found some 
signs of a  large-scale  
anisotropic quasi-periodic structure
detectable  through 6  sectors  
out of a total  of 144 sectors.
These   sectors   are
distinguished  by  large  amplitudes of dominant peaks 
in their  radial   power spectra
at  wavenumbers  $k$  within  
a narrow  interval  of $0.05 < k  <  0.07$~h~Mpc$^{-1}$.
Then, passing from a spherical coordinate system to a Cartesian one, 
we   found   a  special  direction 
such that   the total distribution of LRG projections on it 
contains  a  significant  ($\ga$5$\sigma$)
quasi-periodical component. 
We assume that we are dealing with  a  signature of  
a quasi-regular  structure  
with a characteristic scale   $116 \pm 10$~h$^{-1}$~Mpc.
Our  assumption is  confirmed by 
a preliminary analysis of  the SDSS DR12 data.}

%
{\it  statistical methods;  distances and redshifts of galaxies; 
large-scale structure of Universe} 
%


%
%
\section{Introduction}\label{sec1}

The large-scale distribution of the matter (dark  and  baryonic  substances)  
in the Universe represents a very complex multi-scale structure
(known as {\it {cosmic web}}, 
[1])  
whose origin, evolution, dynamics and 
structural features have been the subject of  extensive  study 
for a few  decades  
(e.g.,  [2]).
The structure represents 
a network of high-density regions formed by galaxy clusters and superclusters, 
walls  and  filaments,  delineating   low-density  
regions  {$-$} 
giant  voids,  occupying the bulk of the space in the Universe  
(for review  see, 
e.g.   [3-5]).

The complex pattern of the cosmic web includes a huge variety of scales, 
ranging from units and tens of megaparsecs up to hundreds of megaparsecs. 
In this regard, it seems important to 
question the highlighted scales of inhomogeneities that appear 
in the largest observable  structures 
and the related question of the possible existence of some  geometric order, 
at least in certain parts of the cosmic web.  
Among the largest  scales,  the most frequently mentioned scales in the literature 
are  $\ga100h^{-1}$~Mpc, 
where 
$h=H_0/100$~km~s$^{-1}$~Mpc$^{-1}$ and $H_0$ is the 
present Hubble constant.
These are, for~instance, the~characteristic scales of large voids  $\sim 100~h^{-1}$~Mpc, 
the spatial scales   (100--110)~$h^{-1}$~Mpc  
corresponding to the Baryon Acoustic Oscillations
(BAO,\ e.g.,~[6-8]
and references therein), 
as well as  somewhat larger scales 
found in ordered quasi-periodic formations.
It is well known that the scale of the BAO is determined by the size
of  the horizon of sound waves   in the recombination epoch   
and manifests itself as the presence of a weak periodic component
in the 3D power spectrum of cosmological inhomogeneities  
(e.g., [9-11]).
 As a consequence of such oscillations in k-space, 
a  significant  bump is registered   
in the spatial 3D correlation function at the
scales  noted above  (e.g., 
[12-16]
and references therein).

On the other hand, 
we have a number of observational evidence 
that some areas of the cosmic web show 
elements of spatial  regularity  at 
scales of the order of (110--140)~$h^{-1}$~Mpc. 
Important  evidence of the existence of some order 
in the spatial distribution of galaxies was  the
detection  of   1D quasi-periodicity  
at a scale $\sim$130~$h^{-1}$~Mpc
found in the pencil-beam surveys 
near both Galactic poles
[17]. 
This result was confirmed  
(but   see  note  in  Section~\ref{sec6}) 
in further studies of pencil-beam distributions of galaxies 
[18-20]. 
This was followed by a series of works, e.g.,~
[21-26],
in which it was shown that the cosmological 
network formed by rich clusters and
superclusters of galaxies and voids between them may show 
traces of a regular spatial (cubic-like or shell-like) structure, 
with characteristic scales  (110--140)~$h^{-1}$~Mpc.

Using a 2D Fourier transform technique (developed in 
[27, 28] 
for clustering of astrophysical objects),
the authors of  [29] 
demonstrated 
the possibility of the existence
of quasi-periodic structures
inside thin slices  
whose centers were directed along 
the  right ascensions  of the  north and south Galactic poles.
In some directions of 
2D  wave vectors  ${\bf k}$, 
significant peaks in 2D power spectra were  detected in selected slices,  
which correspond 
to quasi-periods of order  100~$h^{-1}$~Mpc. 
Somewhat  later,
a new method was proposed in  [30]
for determining the periodicity of
a cubic-like lattice formed by a network  
of superclusters and voids. 
It was shown that a  periodicity with periods
(120--140)~h$^{-1}$~Mpc can be observed  along 
certain directions in~space.

In our previous papers, we  analyzed
radial (shell-like) distributions 
of cosmologically distant matter traced  
by  the luminous red galaxies 
(LRGs;\  
[31]   
hereafter Papers~I) 
or the brightest cluster galaxies 
(BCGs;\  
[32]). 
We treated the radial distributions of matter   
as a sensitive  way  to detect 
possible quasi-periodic 
spatial distributions of cosmological objects. 
When using  the radial  statistics,  the sample is characterized only
by a comoving radius  (light-of-sight distance)
independently of  its  direction  on the sky,
which corresponds to a  complete loss 
of tangential statistical~information. 

Similar approaches were applied, 
for  instance,  
in  [33], 
for searching for
radial shell-like associations of main galaxies 
around central LRGs to visualize the  BAO phenomenon
in the spatial distribution of galaxies,  
or 
in [26]  
for searching  for shell-like structures 
of rich galaxy clusters in
the environment of central~superclusters.

In  the  papers  cited above,   
we found that the radial distribution of 
LRGs  and  BCGs  incorporates a set of
quasi-periodical components relative to the radial
comoving distance.  The~major  scale 
revealed in these  studies  turned out to be
$\sim$$100~h^{-1}$~Mpc.
It was shown, in~particular,  that the existing methods for assessing
the significance of peaks in the power spectrum of radial distributions
(especially in cases with complex behavior of 
smoothed power spectra, so-called   trends in $k$-space ) 
can give a variety of results. 
Therefore, in~[34] 
(hereafter Paper~II), 
a special approach was proposed to assess the significance 
of peaks in  the power spectra of  radial
distributions of objects (galaxies and  clusters),  which 
are subject to clustering at  a variety of  scales. 
This approach systematically
reduces 
(in relation to Paper~I and  [32])
the significance of the peaks 
(up to  $\la3\sigma$) 
in  the  radial power spectra, 
although the peak amplitudes 
may   appear to be quite  large.

The present work  also  refers  to  the  topic 
of  searching  for  quasi-periodic (quasi-ordered) 
structures at  moderate  cosmological redshifts.  
As in  Paper~I,  we  deal with the LRG data by  Kazin~et~al. (2010)
[35]
presented on the World Wide Web 
\footnote{\it {https://cosmo.nyu.edu/~eak306/SDSS-LRG.html}}.
We use their full flux-limited sample DR7-Full  
($0.16 \leq z \leq 0.47,\  -23.2 \leq M_g \leq -21.2$) and two
subsamples with a focus on volume-limited 
regions {$-$} 
DR7-Dim  ($0.16 \leq z \leq 0.36,\ -23.2 \leq M_g \leq - 21.2$) 
and DR7-Bright ($0.16 \leq z \leq 0.44,\ -23.2 \leq  M_g  \leq -21.8$).
DR7-Dim is quasi-volume-limited subsample, while 
DR7-Bright is closest to being volume-limited, 
i.e., the most homogeneous 
of the three  (e.g.,  [36]).  
This  allows  studying 
the variations in the spatial  distribution of LRGs
at different sample homogeneities.

Here, we use data of the mock 
Large Suite of Dark Matter Simulations
(LasDamas) catalog 
(e.g., [35, 37])
and combine these data with the procedure for estimating 
the peak significance, 
proposed in Paper~II.
The procedure is 
based on the exponential distribution of the height 
of random peaks in the power spectra.
The LasDamas catalog was originally  coordinated 
with the data of the SDSS DR7;
therefore, in the present consideration,
we mostly  use 
the same data release
(except  Section~\ref{sec5}).

Based on the SDSS DR7 data  
with redshifts  $0.16 \leq  z  \leq  0.47$,
we  found that within a wide rectangular 
region  in the sky, there are six quite
narrow  restricted areas (sectors) 
characterized by an increase in the amplitudes 
of the dominant peaks  
in the respective radial power spectra. 
The significance evaluation of  these
peaks gives   {$\ga3-5\sigma$}. 
The peaks lie in a narrow range 
of wavenumbers
$0.05<k<0.07~h$~Mpc$^{-1}$  
and  correspond to
quasi-periodical components with rather 
close   periods  ($2\pi/k$). 
However, the~periodicities demonstrate 
markedly different phases and  
the resultant radial power spectra calculated for 
summarized data of all
six sectors  
(as well as for the whole rectangular region)
are essentially smoothed~out.   

On  the other hand,    
specific directions (axes) 
in space were found such  that
the projections  of  
the Cartesian coordinates of  
LRGs,  observed  through  the  six  sectors (windows),
form a one-dimensional (1D) distribution,
which contains a quasi-periodic  component
with  a characteristic scale $116 \pm 10~h^{-1}$~Mpc
at a high level of significance  
({$\ga5\sigma$}). 
Given this, one can imagine that
the quasi-periodic component 
is likely to  represent a set of flat-like
condensations  and rarefactions   
transverse to  a  narrow  beam of axes.
In particular, such a structure  could   give  rise to  
a moderate {$(\la3\sigma)$} 
quasi-periodicity in  the 
radial distribution calculated for  
the whole region under~study.

In Section~\ref{sec2},  we  determine  
basic quantities and definitions used in our analysis
of radial  distributions. 
In Section~\ref{sec3}, we introduce a rectangle region 
in the sky and  find  six sectors  
within this region 
for which the significance of the peaks
in the radial power spectra
at  $0.05<k <0.07~h$~Mpc$^{-1}$  
turns out  to be  relatively high.
In Section~\ref{sec4},  we enter a 
Cartesian coordinate system (CS) and show that 
there is a  narrow beam of axes in the comoving space  
such that the distributions of the coordinate 
projections on these axes
manifest enhanced 
quasi-periodic components. 
In Section~\ref{sec5},  we compare our results  with those
obtained in a similar way but based on the  preliminary analysis  
of SDSS DR12 data.  
Conclusions and discussions 
of the results are  given in Section~\ref{sec6}.

\section{Basic~Definitions}\label{sec2}

The basic value of the power spectra calculations 
is a radial (shell-like) distribution  function $N_R(D)$ 
integrated over angles $\alpha$
(right ascension) and $\delta$ (declination); 
$D = D (z)$ is the line-of-sight comoving distance
between an observer and cosmological objects under study;
$N_R(D)$ d$D$ is the number of objects inside an interval  d$D$. 
The radial comoving distances are calculated 
in a standard way 
(e.g., [38, 39]) 
%
\begin{equation}
D (z_i) = {c \over H_0}\, \int_0^{z_i} 
{1 \over \sqrt{\Omega_{\rm m} (1+z)^3 + 
\Omega_\Lambda}}\ {\rm d}z, 
\label{D}
\end{equation}
where $i$  numerates  redshifts $z_i$
of cosmological objects in a sample, 
$H_0=100~h$~km~s$^{-1}$~Mpc$^{-1}$ is the present Hubble constant, and
$c$ is the speed of light; 
hereafter, 
we use the  same $\Lambda$CDM model
with the relative total density of matter $\Omega_{\rm m}=0.25$ and 
relative   dark energy 
density
$\Omega_{\Lambda}=1-\Omega_{\rm m}=0.75$
as  it is chosen  in  the mock  LasDamas catalogs 
and in  Paper~I.

We use the binning approach 
and  calculate the  so-called 
normalized radial distribution function in the comoving 
CS  as~the  number of redshifts (LRGs) inside concentric (spherical)
non-overlapping bins:
\begin{equation}
{\rm NN}(D_c^l) = 
{N_R (D_c^l) - S
\over \sqrt{S}},
\label{NN}
\end{equation} 
where $D_c^l$ is  the  central  radius of a concentric  bin 
with a width $\Delta_{D} = 10~h^{-1}$~Mpc,
$l=1,2,\dots,{\cal N}_b$ is a numeration of  bins,
$S = \langle N_R \rangle$ is the mean value of  the radial distribution over 
all  bins under  study   
\footnote{Note that instead of the radial distribution  
function $N_R(D)$, one can use a comoving number density 
$n(D)= N_R(D)/{\rm d} V $, where ${\rm d} V$ is a comoving 
differential volume, which is a variation of the conventional
value $n(z)$ 
(e.g.,  [35, 40]). 
In~this case, Equation~(\ref{NN}) 
can be written as \mbox{$NN(D) = (n(D) - \langle n \rangle )/ \sigma (n)$},
where $\sigma (n) = \sigma(N_R)/{\rm d V}$,\  $\sigma$ 
is the mean squared  deviation.}. 
For the majority of  the  distances  $D(z)$  analyzed  hereafter 
(except  DR7-Dim and DR7-Bright data in Section~\ref{sec4},
as well as  the extended interval of DR12 data in Sector~5)
we use a fixed interval $464 \leq D (z) \leq 1274$ for
the redshift region $0.16 \leq z \leq 0.47$, which consists of 
${\cal N}_b = 81$ spherical bins 
\footnote{The bin width $\sim$10~h$^{-1}$~Mpc 
is selected for convenience. It was
specially verified that further results do not depend on bin sizes 
within an interval $\sim$1--10~$h^{-1}$~Mpc,
if we are interested in scales $\sim$100~h$^{-1}$~Mpc.}.

The values of ${\rm NN}(D_c^l)$  allow one to  
calculate the radial  power spectrum
constructed according to the 
definition of  1D power  spectra
(e.g.,  [41,  42])
%
\begin{eqnarray}
\label{PRkm}
P_R (k_m) & = & |F_R^{1{\rm D}} (k_m)|^2 =  
\nonumber    \\
& & \frac{1}{{\cal N}_b}  \left\{ \left[ \sum_{l=1}^{{\cal N}_b}
{\rm NN} (D_c^l)\  \cos(k_m D_c^l) \right]^2   +  \right.
\left.  \left[ \sum_{l=1}^{{\cal  N}_b}  
{\rm NN} (D_c^l)   \sin(k_m D_c^l) \right]^2  \right\},
\end{eqnarray}
where   $ F_R^{1{\rm D}} (k_m) = ({\cal N}_b)^{-1/2}   
\sum_{l=1}^{{\cal N}_b} {\rm NN} (D_c^l)  e^{-i k_m D_c^l}$
is the one-dimensional discrete  Fourier transform, 
$k_m=2\pi  m/L_R$ is a wavenumber
corresponding to an integer harmonic number
$m=1,2,\dots,{\cal M}$,\     ${\cal M}=\lfloor {\cal N}_b/2 \rfloor$ is
a maximal number (the Nyquist number) 
of independent discrete harmonics, $\lfloor x \rfloor$ denotes 
the greatest integer  $\leq$$x$,  $x$ is an arbitrary real (positive) number, and 
$L_R$  is the whole interval 
in the configuration space, i.e.,~the so-called {\it {sampling length}}. 

To assess the significance of the peak amplitudes in the power spectra
of the normalized  {\it {radial}} LRG distributions, we employ 80 ``ns'' (north--south)
realizations of two mock galaxy LasDamas (LD) catalogs, 
``lrgFull-real''  and  ``lrg21p8-real'' 
\footnote{\it  {http://lss.phy.vanderbilt.edu/lasdamas/mocks/gamma}}.
Both  catalogs simulate possible clustering of the LRG distribution 
in accordance with the data obtained by SDSS DR7.   
The first one simulates DR7-Full   and 
DR7-Dim   catalogs  (see Introduction), the~second  {$-$} DR7-Bright.  
Employing  Equations~(\ref{NN}) and  (\ref{PRkm}) 
or their modification
(considered in  Section~\ref{sec4}),
we computed a set  of  power spectra  
for  ${\cal N}_{\rm LD} = 80$  realizations
by considering  each  region 
in the sky  selected~below separately.

When calculating   the  radial power 
spectra $P_{\rm LD}(k)$  for any realization
of  the ``lrgFull-real'' catalog,
we need to carry out  a  scaling (reduction) procedure  as  employed
in Paper~I.  Actually,  radial smooth functions (trends) 
of the LD data  $N_{\rm tr}^{\rm LD} (D_c)$  and 
the complex trend of the LRG sample $N_{\rm tr} (D_c)$
are quite different  
and mutually poorly matched.
Therefore, to~make two types of  the samples more  comparable, 
we   apply  an appropriate   scaling 
\footnote{The divergence of both the trends was discussed
in   [35]; 
the authors  used similar scaling  of the
smoothed LD curve in their   Appendix A.}. 

For this, we  perform   the reduction procedure for  all realizations
of the LD catalog within the  full  available interval $0.16 \leq  z \leq 0.44$  or   
$464 \leq D (z) \leq 1194~h^{-1}$~Mpc\   
using \mbox{a  formula}:  
\begin{equation}
N_{fin}^{\rm LD} (D_c^l) =  N_{in}^{\rm LD} (D_c^l) \cdot  N_{\rm tr}(D_c^l)/ 
N_{\rm tr}^{\rm LD} (D_c^l),
\label{reduc}
\end{equation}
where  index $l$  as in (\ref{NN}) and  (\ref{PRkm}) numerates  bins,  
but with a  slightly 
different bin number  (${\cal N}_b=73$), 
$N_{in}^{\rm LD} (D_c^l)$ and $N_{fin}^{\rm LD} (D_c^l)$
are {\it initial} and {\it {final}} radial distributions of  mock galaxies 
over all investigated  bins,  
$N_{\rm tr}^{\rm LD} (D_c^l)$ is a  trend calculated 
for each mock realization, 
$N_{\rm tr}(D_c^l)$  is  a  trend of the radial distribution calculated 
for a  sample of LRGs, and both  are obtained   
employing  the least-square method with  a set of parabolas.  
Using   Equation~(\ref{NN}),
we determine the normalized radial distribution  NN$^{\rm LD} (D_c^l)$,
where  $N_{fin}^{\rm LD} (D_c^l)$   and  the  mean 
$S^{\rm LD}$ (over the whole indicated interval) 
stand for  $N_R (D_c)$ and $S$, respectively.
It is worth emphasizing   that 
all calculations of the power spectra 
(\ref{PRkm})  are  carried out  
in a uniform way,  avoiding the concept of  a  trend.  
This guarantees
an  undistorted representation 
of all scales in the power spectra.

Similar to  [43],  
we  calculate   
a power spectrum $\langle P_{\rm LD} (k) \rangle$ averaged  over all 
80   radial  spectra  $ P_{\rm LD}^n  (k)$ %
\footnote{Here, we mean  so-called  {\it  {ensemble  averaging}}.
However, as~it  is shown in Paper~II,  the {\it  {ensemble averaging}}  
of the radial power spectra  is  equal to  an averaging  
over  many  power spectra calculated for numerous radial distributions
built  relative to different centers, 
i.e.,  so-called   {\it  volume averaging}.} 
and  construct  a corresponding  covariance  matrix
\begin{equation}
C_{i, j} = \frac{\sum_{n=1}^{{\cal N}_{\rm LD}}\  
[\langle P_{\rm LD} (k_i) \rangle - P_{\rm LD}^n (k_i)]
[\langle P_{\rm LD} (k_j) \rangle - P_{\rm LD}^n (k_j)] } 
{{\cal N}_{\rm LD} - 1},
\label{Cij}
\end{equation}
where index $n$ numerates  spectra of different LD realizations,
and $i$ and $j$ run over different  harmonic numbers  $m$ in Equation~(\ref{PRkm}).

As the next step we produce fitting of  the 
average radial  power spectrum $\langle P_{\rm LD} (k) \rangle$
by  a smooth  model  function $f (k)$,  which  is designed as 
\begin{equation}
f (k) = f_{\rm CDM} (k) + 1,
\label{f}
\end{equation}
here,  $f_{\rm CDM}(k)$   
is a 3D power spectrum of the cold dark matter (CDM) density 
averaged over all directions in $k$-space
(e.g.,  [44])
\begin{equation}
f_{\rm CDM} (k) = A  \cdot q\,  {\rm T}^2(q),
\label{fCDM}
\end{equation}
$A$ is a normalizing constant to be found,
and $q$ is a dimensionless 
variable 
determined according to 
[45]  
as 
\begin{equation}
q=\frac{k/({\rm Mpc}^{-1}\ h)}{\Omega_{\rm m}  h  
\exp[-\Omega_{\rm b}(1 + \sqrt{2 h}/\Omega_{\rm m})]},
\label{q}
\end{equation}
where  $k=|{\bf k}|$,   $\Omega_{\rm m}$ is introduced above,\ 
$\Omega_{\rm b}=0.04$ is the relative density of baryons  
(and $  h=0.7$),\
and ${\rm T}(q)$ is a transfer function: 
\begin{eqnarray}
\label{Tq}
& & {\rm T}(q)={\ln (1+2.34 q) \over 2.34 q} \times  \\
\nonumber    
& & [1+3.89 q + (16.1 q)^2 + 
                         (5.46 q)^3 + (6.71 q)^4]^{-1/4}.
\end{eqnarray}

The second term 
 ``1''  on the right-hand side of  Equation~(\ref{f})
stands for so-called ``shot noise'' 
(e.g., \cite{43}), 
which dominates at small scales (large $k$)
and takes into account  additional 
random (Poisson) distribution of point-like~objects.

It was shown in Paper~II 
by numerical calculations
that  $ \langle P_R (k) \rangle = P_{3D} (k) $, 
where $P_{3D} (k)$ is the 3D power spectrum averaged over directions of $\vec{k}$. 
Considering this,  we assume that 
the  average radial  power spectrum $\langle P_{\rm LD} (k) \rangle$  
({\it ensemble averaging})
provides a good approximation for the average radial power 
spectrum of the {\it real} sample of
LRGs  $\langle P_R (k) \rangle$, which in principle   
could be  calculated  as  {\it volume averaging}.
Therefore,   we  use  the  equality  
$\langle P_R (k) \rangle \simeq   \langle P_{\rm LD} (k) \rangle $
in our  assessments~below.  

Then  we can  employ   
the  smooth  function $f(k)$   (see Equation~(\ref{f}))
as  an approximate  substitute of 
$\langle P_{\rm LD} (k) \rangle$. 
To  describe the fit quantitatively,   
we introduce the maximum likelihood function
\begin{eqnarray}
\label{L}
& & L   \propto  \exp[- \frac{1}{2} \cdot \chi^2(A)];   \\
\nonumber  
& & \chi^2(A)  =    \\
\nonumber  
& & [ \langle P_{\rm LD} (k) \rangle - f(A, k)]^T \cdot \hat{C}^{-1} \cdot
[ \langle P_{\rm LD} (k) \rangle - f(A, k)],  
\end{eqnarray}
where the upper index $T$  means transposed matrix, and $\hat{C}^{-1}$ is the inverse matrix 
with respect to  $\hat{C}$  given in (\ref{Cij}).  
Varying the constant  $A$  in Equation~(\ref{fCDM}),  one can  find  the  best  fit  at a minimal
value of  $\chi^2$.  

It was also verified  numerically in  Paper~II  
for a set  of
simulated radial  power spectra $P_R(k)$ that
the cumulative  probability  function of random
peak  amplitudes  $P_k$ 
at any  $k_{\rm max}$ (a  central wavenumber  of  a  peak) 
integrated  over all values  lower than a fixed
value  $P^*_k$  can be expressed as  
(see also, e.g.,~
[42, 46])
%
\begin{equation}
{\cal F}(P_k < P^*_k,\  \lambda)  = 1 - \exp(-\lambda\ \cdot P^*_k )
\, \, \, \, {\rm at} \, \, \, \,  P^*_k  \geq  0,
\label{calF}
\end{equation}
where $\lambda = \lambda(k)$ 
is a parameter of the exponential distribution
determined by a reciprocal  
mean  (mathematical expectation)
peak  amplitude  M$[P_k] = \langle P_R (k) \rangle$,
i.e., $\lambda (k) = \langle P_R (k) \rangle^{-1} \simeq  f^{-1} (k)$.
In this  double equality, 
we  replace  $\langle P_R (k) \rangle$  
by the value  $\langle P_{\rm LD} (k) \rangle$ 
and, in~turn,  by~the function  $f(k)$.
This estimation is valid for a single independent
peak at arbitrary m and yields the probability of
pure noise, generating a power $P(m)$ less than the given level~$P$.

Let us emphasize  also  that 
the difference between Equation~(13)  
of  
[42]  
or  Equation~(7)
of  
[46]
and  Equation~(\ref{calF}) is a  constant
parameter $\lambda$ of the exponential distributions 
in the  cited  papers,  while 
we consider a variable $\lambda (k)$ in the present study 
\footnote{In fact, the~approach of 
[42] and  [46] 
is true in many cases, e.g.,~for radial distributions of absorption 
systems, as~it will be shown in future work.}.

Equation~(\ref{calF}) allows one to build 
fixed confidence probabilities 
for various  $k$ and  connect  them 
in a  single smooth curve 
to outline  an  appropriate  significance level.
The curves obtained in this way can be used as a measure of the significance 
of  separate independent  peaks 
in the power spectra of real LRG samples.
In Figures~\ref{rectPRk}--\ref{Dim-BrightPXk},
dashed lines show  two   levels of significance  ($3\sigma$ and $4\sigma$)
calculated   using data of  all  80 LD  realizations  
by the procedure   described above  
with the  respective  (quasi-Gaussian)  probabilities:
$3\sigma$  {$-$} 
$0.998$, $4\sigma$  {$-$} $0.999936$.
In contrast, the~significance levels  $5\sigma$ (probability
$5\sigma$  {$-$} $0.9999994$) 
are also  shown in Figures~\ref{BrPRk}--\ref{Dim-BrightPXk}
as narrow bands  corresponding  
to values  $A$ (Equation~(\ref{fCDM})) obtained in a similar way but 
within an error interval
$\pm  1\sigma$.     

\section{Radial Distributions in Rectangle Region and~Sectors}\label{sec3}
\begin{figure}[t]     
\begin{center}
\includegraphics[width=0.44\textwidth]{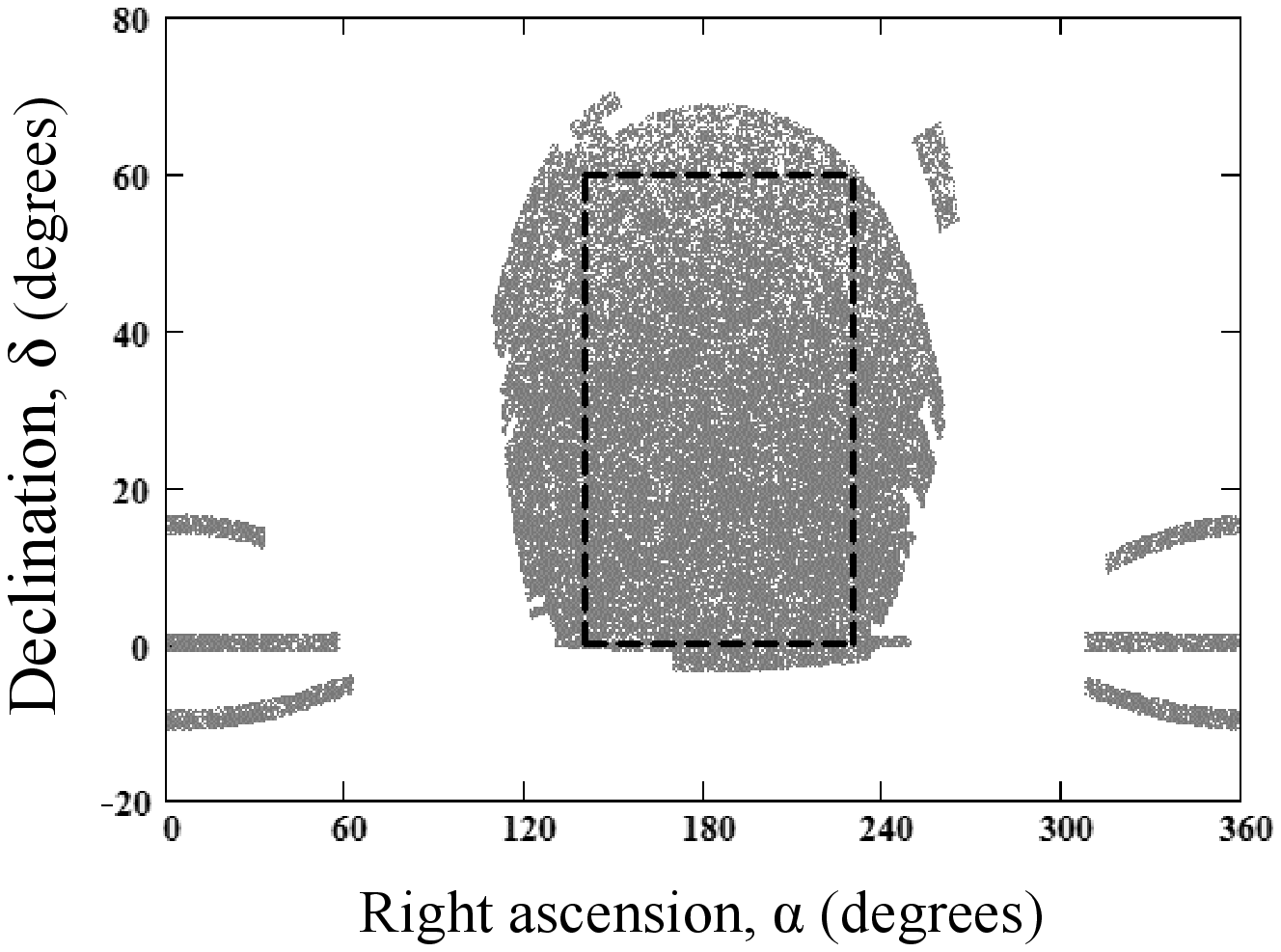}
\hspace{2mm}
\includegraphics[width=0.47\textwidth]{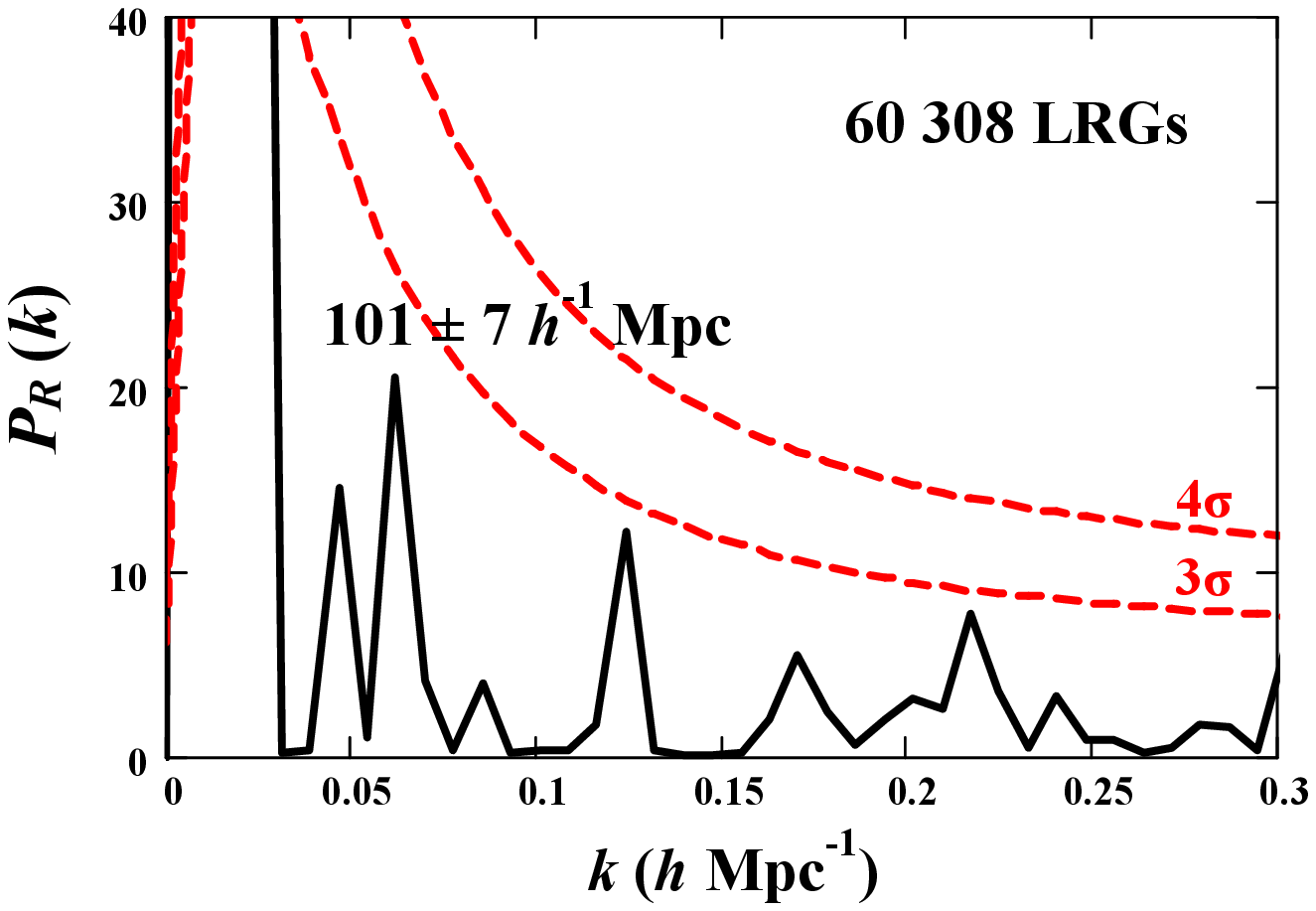}
\caption{
{\footnotesize
(\textbf{Left  panel}): 
Angular distribution of LRGs over the sky from the  SDSS DR7 data 
in the equatorial coordinate system; all grey colored regions 
comprise the  DR7-Full sample  in  [35];
dashed lines delimit the rectangular region of our statistical 
consideration with $\alpha = 140^\circ - 230^\circ$ and
$\delta = 0^\circ - 60^\circ$,\  $\alpha$ 
and $\delta$ are  defined about the axes,  $\alpha$ is shown
in a nonstandard way: east  to  right, west  to  left. 
(\textbf{Right  panel}): 
Radial power spectrum $P_R(k)$ 
solid line calculated according to 
Equation~\protect{(\ref{PRkm})} for all LRGs located in 
the rectangular region 
within the redshift
interval $0.16 \leq z \leq 0.47$\  or 
distance interval
$464 \leq D (z)\leq 1274~h^{-1}$~Mpc.
The significance levels  $3\sigma$  and  $4\sigma$ (dashed lines)
are calculated using the mock LasDamas catalog   and
the exponential probability  function \protect{(\ref{calF})}
for random  spectral peak amplitudes  (see text) } }
\label{rectPRk}  
\end{center}
\end{figure}

\begin{figure}[t]    
\begin{center}
%
\includegraphics[width=0.47\textwidth]{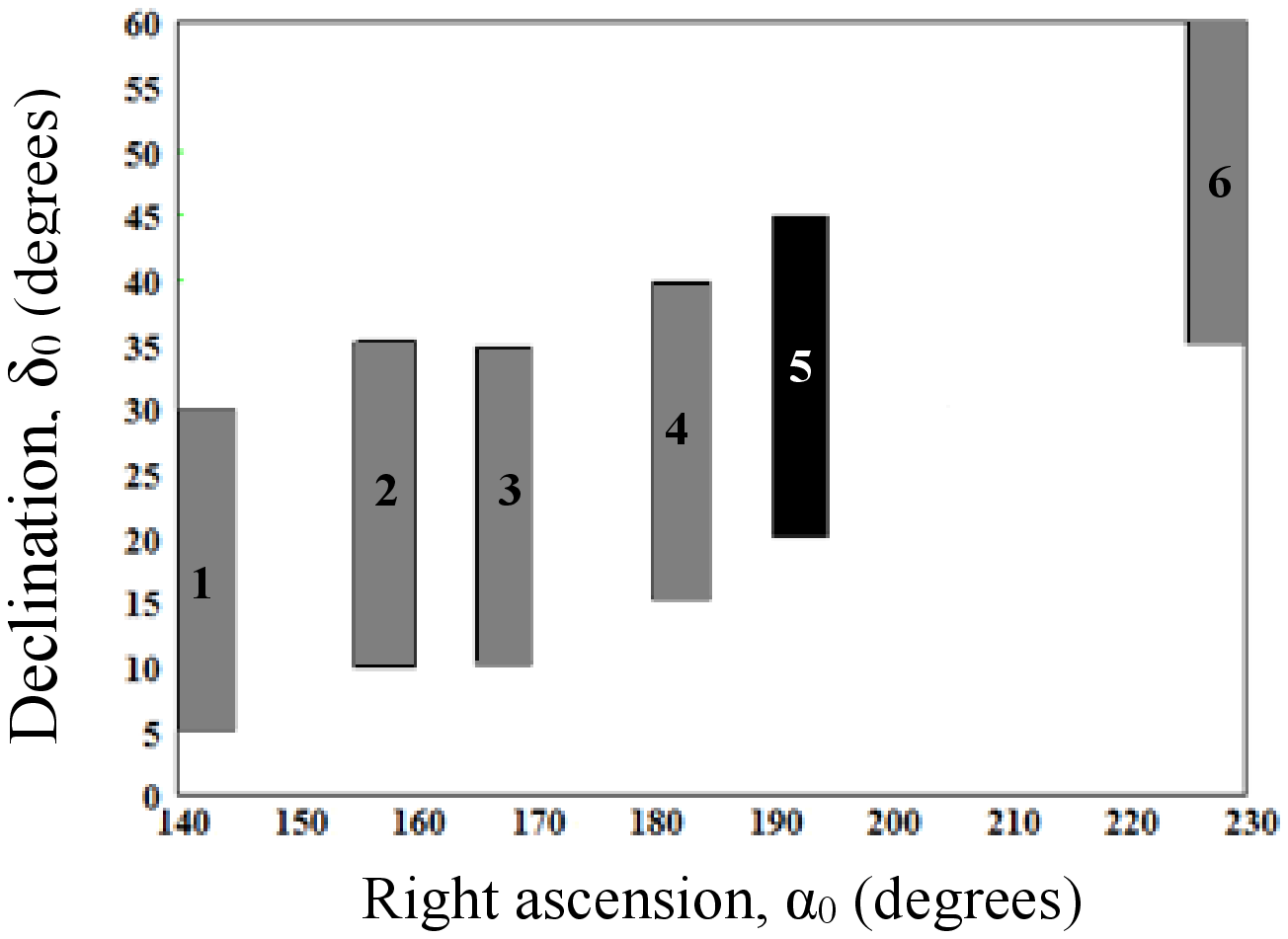}
\hspace{2mm}
\includegraphics[width=0.41\textwidth]{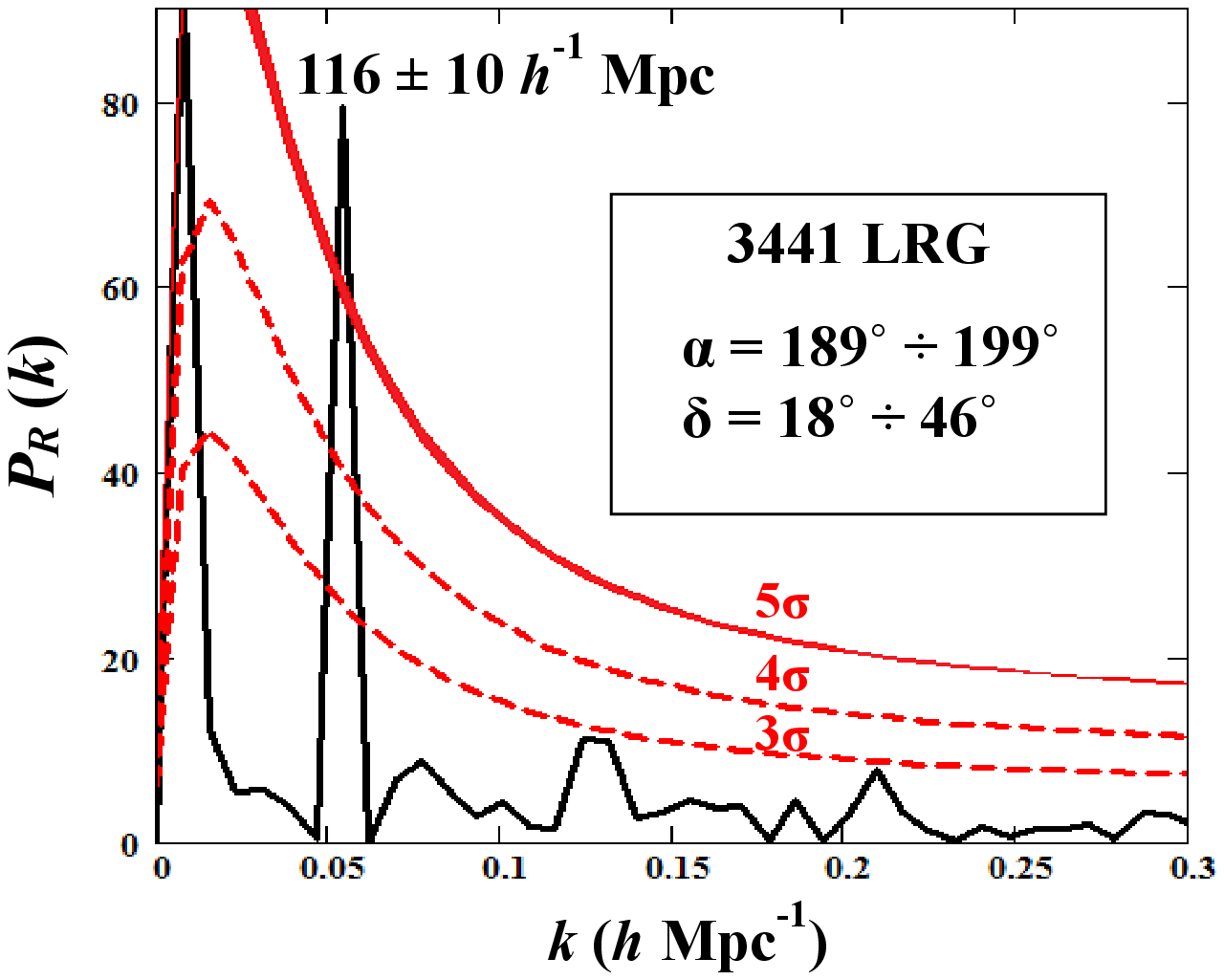}
\caption{
{\footnotesize
(\textbf{Left  panel}): 
A total of 6  out of  144  sectors  
covering the entire 
rectangular area, 
which stand out
due to their high peak amplitudes  in  the
radial power spectra at $k_{\rm max}$, 
lying in  a  range
$0.05 \leq  k_{\rm max}  \leq 0.07~h$~Mpc$^{-1}$.
(\textbf{Right  panel}): 
Radial power spectrum $P_R(k)$ 
(solid line) calculated using 
Equation~\protect{(\ref{PRkm})} for the black sector (No  5) 
in the  left  panel;   dashed lines  $-$\
the significance levels $3\sigma$, $4\sigma$
and a narrow  band  $-$\ $5\sigma$ 
calculated  using  the mock LasDamas catalog (see text);\
intervals of $\alpha$ and $\beta$ 
as well as the sample size are indicated in the insert } }
\label{BrPRk} 
\end{center}
\end{figure}

In this section, we  only use  the  DR7-Full  sample 
as it contains the largest amount of statistical data.
The  SDSS DR7 LRG regions of the sky in the equatorial coordinates 
are shown  in the left  panel of  Figure~\ref{rectPRk}.
We restrict ourselves by considering
a rectangle region highlighted in the left  panel
to avoid the possible effects of  irregular edges 
of the central domain.
In such a way, we choose the intervals of right ascension 
$140^\circ  \leq \alpha \leq 230^\circ$ and declination $0 \leq \delta \leq 60^\circ$.
The sample contains  60,308 LRGs  observed   within the redshift  interval   
indicated above.   Note that in the left  panel of  Figure~\ref{rectPRk}, as  
in the left  panels of the  following three
Figures~\ref{BrPRk}--\ref{rectPXk}, 
the  right ascension  $\alpha$  is shown  in a nonstandard way:
east to  right, west to~left. 
 
The  right  panel of  Figure~\ref{rectPRk} represents the radial power spectrum 
calculated  with the use of Equation~(\ref{PRkm}) at  $0 < k  \leq 0.3$ for 
the entire  rectangle region in the sky. 
The  dominant peaks at $k > 0.04~h$~Mpc$^{-1}$
correspond to  $k_{\rm max}=0.062~h$~Mpc$^{-1}$  
or  spatial comoving scale  $(101 \pm 7)~h^{-1}$~Mpc
and have  
quite large amplitude (about 20).
However, 
the present  evaluation on the base of  the mock  LasDamas catalog
turns out to be  noticeably less than 3$\sigma$.
It  is not  serious enough to discuss 
the quasi-periodicity of the radial LRG distribution. 
Note also  that very large values of $P_R (k)$  
at   the smallest values of $k$ (largest scales), 
$ k < 0.04~h$~Mpc$^{-1}$, 
are   associated with a large-scale trend  $N_{\rm tr}(D)$,  
i.e.,~with  the smoothed part of the total radial distribution function
$N_R (D)$,  and~can be~ignored.   

\unskip
Let us   produce  the next step that takes us beyond purely  radial distributions.
We   scan the entire  rectangle region  by a  trial sector with angular dimensions
$5^\circ \times 25^\circ$  along right ascension and declination, respectively.   
At first,  we  build  the radial distribution of LRGs precisely within the trial  sector
located in the angle range $\alpha = 140^\circ \div 145^\circ$ and 
$\delta = 0^\circ  \div  25^\circ$, i.e.,~in the left  lower  corner  of the rectangle
region on the left  panel of Figure~\ref{BrPRk}.  
Using Equation~(\ref{PRkm}), we calculate the 1D
power spectrum $P_R (k)$ for the radial  distribution of 
those LRGs, which were  observed  through  this  
sector (as through a window) in the sky
and strictly  limited  by  the same  redshift   or distance
intervals, as indicated  in Section~\ref{sec2}.
Then we consequently shift
this trial sector along the axes  of right ascension  or  declination by five degrees
and calculate the appropriate radial power spectra. In~such a way,  we consider 144
radial power spectra $P_R (k)$,   which are  
non-overlapping along the horizontal 
axis but  overlapping along the vertical~one.

When analyzing the obtained spectra, 
we restrict ourselves to  an  interval 
$0.05 < k  <  0.07$ and exploit the same procedure of significance
assessment as it is described  in Section~\ref{sec2}  using the data of
the mock LasDamas catalog.  On~the basis of these data, one can  
build  the radial distributions and calculate power spectra within all
of the outlined  angular  sectors. 
Among  144 sectors,  we select only six,  
in which the significance  of the peak amplitude  exceeds  $3\sigma$,  and
vary  the angular boundaries  of these sectors  with a step $1^\circ$
to achieve maximum peak amplitudes. 
In this way, the  appropriate angular dimensions  
of the six  sectors were found to preserve the  rectangular shape.
After that,  the 1D Fourier transform 
of the radial distributions of LRGs  
(already at $0 < k \leq 0.3$) 
within  their   angular
boundaries  were produced. 
This  allows us to calculate the scales and phases 
of quasi-periodic components in  the  selected  cases. 

\unskip

The  six  selected   sectors are represented 
in the left  panel of Figure~\ref{BrPRk},
and their  characteristics   are shown  in Table~\ref{tab:sectors}.  The~data include
the  sector   numbers,  the~boundaries of angular variables, number of   sampled  LRGs,
quasi-periods  $\Delta D_c$  with  errors determined as  HWHM 
of the main spectral peaks  and  significance of the peaks.  
%
\begin{table}[t]
\begin{center}
\caption{
{\footnotesize
Data on  six selected sectors with relatively high 
significance (Sign.) of the peaks   in   radial power spectra} }
\vspace{0.3cm}
\begin{tabular}{cccccc}
\hline 
{\footnotesize {No}} & {\footnotesize{$\alpha$}}  & {\footnotesize{$\delta$}} &  
{\footnotesize{Number LRGs}}  & {\footnotesize{Periods Mpc/$h$}}   & 
{\footnotesize{Sign }}   \\
\hline
1  & $140^\circ$--$146^\circ$ & $5^\circ$--$31^\circ$ & 2098 & $101 \pm 7$ & $>$$3\sigma$  \\
2  & $154^\circ$--$ 160^\circ$ & $8^\circ$--$39^\circ$ & 2434 & $116 \pm 10$ & $>$$3\sigma$  \\
3  & $162^\circ$--$ 174^\circ$ & $10^\circ$--$35^\circ$ & 3629 & $116 \pm 10$ & $>$$5\sigma$  \\
4  & $182^\circ$--$185^\circ$ & $15^\circ$--$41^\circ$ & 1121 & $116 \pm 10$ & $>$$3\sigma$   \\
5  & $189^\circ$--$199^\circ$ & $18^\circ$--$46^\circ$ & 3441 & $116 \pm 10$ & $>$$5\sigma$   \\
6  & $ 222^\circ$--$230^\circ$ & $36^\circ$--$60^\circ$ & 1634 & $116 \pm 10$ & $>$$3\sigma$  \\
\hline
\end{tabular}
\label{tab:sectors}
\end{center}
\end{table}
\unskip
Only  sector  No~1  in Figure~{\ref{BrPRk}} 
contains a quasi-periodical component
with period  $\Delta D_c = 101 \pm 7~h^{-1}$~Mpc \
($k_{\rm max}=0.062 \pm 0.005~h$~Mpc$^{-1}$)
at   significance 
$>$$3 \sigma$. The~remaining five  demonstrate significant peaks
at  $k_{\rm max}=0.054 \pm 0.005~h$~Mpc$^{-1}$,  
corresponding to quasi-periodicity 
with  a scale  $\Delta D_c = 116 \pm 10~h^{-1}$~Mpc.

The   right  panel in  Figure~{\ref{BrPRk}}  shows  the radial power spectrum
calculated  using the  data of 
sector  No~5 (black one in Figure~\ref{BrPRk}). 
Note  that  the sector 
contains the north  Galactic  pole ($\alpha_{\rm np}=192.86^\circ$ 
and $\delta_{\rm np}=27.13^\circ$;  see discussion in 
Section~\ref{sec6}).  In~this case,  we obtain  the most 
prominent peak among all six sectors  at  the same $k_{\rm max}$ 
as the other five. 
The  sample  size   is  3441 LRGs.
The  significance levels  (also shown in the  right  panel) 
are  estimated using the LasDamas catalog 
within  the same sector on the sky.
The dashed lines plot  
the significance levels  $3 \sigma$ and $4 \sigma$, while the 
narrow band  plots  the level  $5 \sigma$
also calculated using  (\ref{calF}) but 
for  $A$  (see  Equation~(\ref{fCDM}))  lying  within   
a  $1 \sigma$ error  interval  
(in this case $A \simeq 160 \pm 4$).

A single peak at  $k =0.054~h~{\rm Mpc}^{-1}$ 
markedly  exceeding  $5 \sigma$
is clearly visible
in the power spectrum. 
This can serve as an additional 
justification  for  using  Equation~(\ref{calF})
to estimate  the significance of separate peaks 
at different $ k $  as  a  result of independent 
random fluctuations.
Actually, smooth lines representing the significance levels 
on the right panel of Figure~\ref{BrPRk} and in the following figures 
are  the locus of single peaks at  fixed significance, calculated 
employing  LasDamas data and the formulas 
(\ref{f})--(\ref{calF})  or  their modifications 
(see Sections~\ref{sec4} and \ref{sec5}). 
On the other hand, it can be shown 
following 
[42, 46] as well as
[47, 48]  (and references therein)
that 
the probability  of occurrence of
any number of random independent peaks
for different $ k $
leads to levels of significance not too different 
from that  calculated with  Equation~(\ref{calF}) \footnote{
In fact, 
one  can consider a set of many independent 
wavenumbers $k_m$;
$m = 1, 2,\dots, {\cal M}_k$ , where   
${\cal N}_b =81$  and   ${\cal M}_k=40$ 
(see text under Equation~(\ref{PRkm})),
and  treat any of the  spectral   peaks $P(k_m)$ 
as a result of Gaussian noise.
Then, one can estimate the so-called false alarm probability,
Pr$(P_{\rm max} \geq  P^*) = {\cal P}_0  = 1 - \beta^{{\cal M}_k}$, 
where $\beta \equiv {\cal F}(P_k  \leq  P^*_k)$  
is defined  in  Equation~(\ref{calF}),
i.e., ${\cal P}_0$ is  probability of at least one  
 of many possible peaks
$P_{\rm max}={\rm p}_0$
being equal to
(or above) a maximal level $P^*$.  
Using the formula  
${\rm p}_0 = - \lambda^{-1}\ \ln[1- (1- {\cal P}_0)^{1/{\cal M}_k}  ]$
(see [42, 46])   
at   $k = k_{\rm max} = 0.054$
one can obtain  that  the confidence  level  
$(1 - {\cal P}_0) = 0.9999994$ (significance $5 \sigma$ ) 
corresponds
to ${\rm p}_0 \simeq 74.0$. 
Thus, the calculated  value of  the peak amplitude
in the right panel of Figure~\ref{BrPRk}
$P_{\rm max} = 79.3$    lies
noticeably higher than  ${\rm p}_0$. }.

On the other hand, 
the power spectrum of the total normalized distribution 
of LRGs in all six selected sectors over the entire interval 
$0.04 \leq  k  \leq 0.3~h$~Mpc$^{-1}$ 
does not contain significant peaks. 
This  is  similar to the spectrum obtained  for the  whole rectangle 
region and is plotted  in the  right  panel  of   Figure~\ref{rectPRk}. 
The small significance of the period 116~$h^{-1}$~Mpc
is a consequence of the fact 
that  these  harmonics  in  the selected  five sectors 
have different phases and mutually  extinguish
each~other.

Random  appearance and  disappearance 
of the quasi-periodicity during the rotation 
of  an observation  field  from sector to sector
on  the celestial sphere, as~well as  random
phase shifts between the selected sectors, 
might  indicate  the existence of
a sparse  and  ragged  spatial structure 
at   large  cosmological distances. 
It may  be assumed that  radial distributions
within  wide angular regions 
are only capable  of tracing such a structure   indirectly.
Below we develop a different approach for  searching and  analyzing 
such  a  possible structure and   assessing  its significance.

\section{Cartesian Coordinate System.  Preferred~Direction}\label{sec4}

In this section, we refer to all three samples of LRGs 
presented  in Section~\ref{sec1},
DR7-Full, DR7-Dim and DR7-Bright, selected and described  in  [35]
(see also    [40]).

Let us move from spherical coordinates  characterizing  LRGs in 
Sections~\ref{sec2}  and  \ref{sec3} 
to the  distribution of the LRGs  in Cartesian  CS:
\begin{eqnarray}
\label{XYZ}
&  & X_i = D (z_i) \sin(90^\circ - \delta_i) \cos \alpha_i       \\
\nonumber  
&  & Y_i = D (z_i)  \sin(90^\circ - \delta_i) \sin \alpha_i          \\
\nonumber  
&  &  Z_i = D (z_i) \cos(90^\circ - \delta_i),
\end{eqnarray}
where $D (z_i)$ is the radial comoving distance of  
$i$-th  LRG with redshift  $z_i$,
$\alpha_i$ {$-$} its right ascension and $\delta_i$ {$-$} declination;
in both the coordinate systems,  an observer is at the zero~point. 

Following  the definitions  of
Section~\ref{sec2}, we use the binning approach 
along the axis $X$,  
and similar to calculations of $N_R(D_c)$, 
we can calculate
a distribution  $N_X (X_c)$, where $X_c$  
is  a  central  point of  a  bin, and $\Delta_X$ is  its  width. 
For the sample DR7-Full, we fix the same analyzed range  
$464 \leq  X  \leq  1274~h^{-1}$~Mpc 
containing the same ${\cal N}_b = 81$ independent bins with  
a width $\Delta_X = 10~h^{-1}$~Mpc as it is used   in\   (\ref{NN})
for $D(z)$  (the intervals of
DR7-Dim and DR7-Bright are considered below).  

By analogy with Equation~(\ref{NN}),  we calculate 
the normalized  1D distribution 
along  an  axis $X$
\begin{equation}
{\rm NN}_X (X_c^l) = {N_X (X_c^l) - S_X
\over \sqrt{S_X}},
\label{NNX}
\end{equation} 
where  $l=1,2,\dots,{\cal N}_b$ is  also  the  numeration of  bins, and
$S_X$ is  a mean value of  the 1D  distribution $N_X (X_c^l)$ over all 
bins.  

Using Equations~(\ref{PRkm}) and (\ref{NNX}),  one can calculate
the 1D  power spectrum $P_X (k_m)$  
replacing in   (\ref{PRkm})  
 $D_c^l$   by   $X_c^l$ and 
${\rm NN}(D_c^l)$   by   ${\rm NN}_X (X_c^l)$;\ 
in this case,    $k_m=2\pi  m/L_X$ is a wavenumber,
$m=1,2,\dots,{\cal M}$ is a harmonic number,     
${\cal M}=\lfloor {\cal N}_b/2 \rfloor$ is
a maximal number (the Nyquist number), and
$L_X = 810~h^{-1}$~Mpc  is the whole interval along the axis $X$
(sampling length). 

Then, we  rotate the coordinate axes  $XYZ$ at certain Euler angles
so that the new axis $X'$%
\footnote{Hereafter, denotation $X'$ instead 
of general denotation $X$ indicates the  axis
rotating together with the rotation of the CS.}  
would be  oriented  in a certain  direction 
($\alpha'$ and $\delta'$)  relative to the  initial Equatorial CS.  
Performing a sequence of such rotations, 
we search for the  $X_0$-axis  along which  
the 1D power spectrum  calculated
for all  six  sectors in total 
displays  the most  significant peak 
at  a  scale   $\sim 116~h^{-1}$~Mpc.

To control  the uniformity of statistics  for different 
directions of $X$,
we fix the same  boundaries of the rotated axes, e.g.,~
$464  \leq  X  \leq  1274~h^{-1}$~Mpc for the sample DR7-Full.
This  condition strongly  limits the area  of 
analyzed directions\  $160^\circ \leq \alpha \leq 200^\circ$ and 
\mbox{$20^\circ \leq \delta \leq 40^\circ$} 
inside the rectangle. The~same angular limits are set 
for the other two samples  to ensure the same conditions 
in all cases under study.  
Employing the modification
of  \mbox{Equation~(\ref{PRkm})} described above,  
we calculate the 1D  Fourier transform $F_X (k_m)$ and 
the power spectra  $P_X (k_m)$
for each direction of $X$. 

Actually, we deal with
a  discrete analog of  so-called 3D  Radon transform 
(e.g., [49])
applied to  selected 
data, i.e.,~we  summarize all  the  points  from a subsample 
whose projections fall into each bin given along $X$. 
Thereafter, we  exploit the two main properties of the Radon transform
(i) {\it translation invariance}, which allows one  to transfer the projections 
of  the Cartesian  galaxy coordinates on the given axis  $X$
to another axis  $\hat{X}$  parallel to the original  one,
(ii)  {\it linearity}, which allows one  to  
summarize the  projections obtained for individual sectors 
in the sky into the total sum of  projections
to get  a single Radon transform 
for  the entire  sample.

We  start with an orientation of  the  $X$-axis along a direction
with coordinates $\alpha=160^\circ$ and $\delta = 20^\circ$
(lower left corner of the indicated region) and  
rotate   the axis $X'$ sequentially,
shifting the right ascension or declination
with a step $1^\circ$. Note that such  rotations 
of  the moving CS require only two
Euler angles, $\alpha_{\rm Eu} = \Delta \alpha$ and 
$\beta_{\rm Eu} = \Delta \delta$, where  $\Delta \alpha$
and  $\Delta \delta$ are respective rotation angles.
As a result,   one  can find  
an axis  $X_0$ with Equatorial 
coordinates $\alpha_0$  and  $\delta_0$ in the sky 
along which the 1D distribution 
of the Cartesian coordinate projections  shows
the maximum peak amplitude  at  
$k$$\sim$0.05--0.06~$h$~Mpc$^{-1}$. 

\begin{table}[t]
\begin{center}
\caption{\footnotesize{SDSS LRG (DR7) and (DR12) statistics }}
\vspace{0.3cm}
\begin{tabular}{cccccc}
\hline 
{\footnotesize {Sample}}   & {\footnotesize{Redshift}}  & {\footnotesize{ $X$ $^{a}$ \   Mpc~$h^{-1}$}} &  
{\footnotesize{Number LRGs}}  & {\footnotesize{$ \alpha_0$ $^{b}$}} & {\footnotesize{$\delta_0$ $^{b}$ }} \\
\hline
{\footnotesize{DR7-Full}}  $^{c}$ & $0.16 \leq z  \leq  0.47 $ &  $ 464 \leq X  \leq 1274$ & 13 872 &
 $176^\circ$ & $ 24^\circ$  \\
{\footnotesize{DR7-Dim}}  $^{c}$  & $0.16 \leq z \leq  0.36 $ & $464 \leq X \leq 994 $ & 8606 &
$177^\circ$ & $22^\circ$ \\
{\footnotesize{DR7-Bright}}  $^{c}$ & $ 0.16 \leq z  \leq 0.44 $ &   424 $^{f} \leq X  \leq 1114$  & 4185  &  
$177^\circ $ & $ 23^\circ$  \\
{\footnotesize{DR12-LOWZ}}  $^{c,\ d}$  & $0.16 \leq  z \leq  0.47$ & $464 \leq X  \leq 1274 $   &  38 880 &  
$175^\circ $ & $25^\circ$ \\
{\footnotesize{DR12-SMASSLOWZE3}} $^{c,\ d}$ & $0.16 \leq  z \leq  0.72$  & $464 \leq  X  \leq 1844$   &  106 136   & 
$176^\circ$ & $27^\circ$    \\
{\footnotesize{DR7-Full}} $^{e}$ & $0.16 \leq  z \leq 0.47 $ &  $464 \leq  X  \leq 1274$ & 
57 099 & $175^\circ$ & $27^\circ$  \\
\hline
\end{tabular}
%
\begin{tabular}{l}
{\footnotesize
$^{a}$  Intervals of  galaxy  Cartesian   coordinate projections on  any  $X$-axis  (see text). } \\
{\footnotesize
$^{b}$  Equatorial coordinates  of  $X_0$-axes selected  for each  sample  (see text). } \\
{\footnotesize
$^{c}$  The data refer to the six sectors in the sky shown in Figure~\protect{\ref{BrPRk}}. } \\
{\footnotesize
$^{d}$  Two samples from  the DR12 data (see Section~\ref{sec5}). } \\
{\footnotesize
$^{e}$  The data refer to  LRGs observed  in the whole rectangular region } \\
{\footnotesize
\quad   shown  in Figure~\protect{\ref{rectPRk}}. } \\
{\footnotesize
$^{f}$  Special case: $X \geq 424~h^{-1}$~Mpc  at $z \geq 0.16$ (see text). }
\end{tabular}
\label{tab:samples}
\end{center}
\end{table}
\unskip

Table~\ref{tab:samples} 
gives  the main characteristics of the subsamples
used  to find  directions $X_0$ 
and to estimate  the significance level  
of the main peak in each case.
For all three subsamples, we perform  rotations 
of the $X$-axis at fixed boundaries, $X_{\rm min}$ and
$X_{\rm max}$ (different for each sample),  
providing  uniformity of 
statistical  conditions  in  different directions \footnote{
Note that  the relationship between the boundaries of $z$ and $X$ is ambiguous.
Using this ambiguity, we shift the lower bound
$X_{\rm min}$  of  the sample DR7-Bright 
to $424~h^{-1}$~Mpc  relative to 
$464~h^{-1}$~Mpc  fixed  in  the other  two  cases
of DR7 data, 
thereby slightly expanding the interval $X$ 
with the same $z \geq 0.16$.  On~the other hand, 
the upper  boundaries of DR7-Dim and DR7-Bright are also shifted
relative to the $X_{\rm max} = 1274~h^{-1}$~Mpc  accepted
for DR7-Full because these subsamples 
correspond to  lower $z_{\rm max}$, i.e.,~
$\leq$0.36  and $\leq$0.44,
respectively}.

\begin{figure}[t]
\begin{center}
\includegraphics[width=0.60\textwidth]{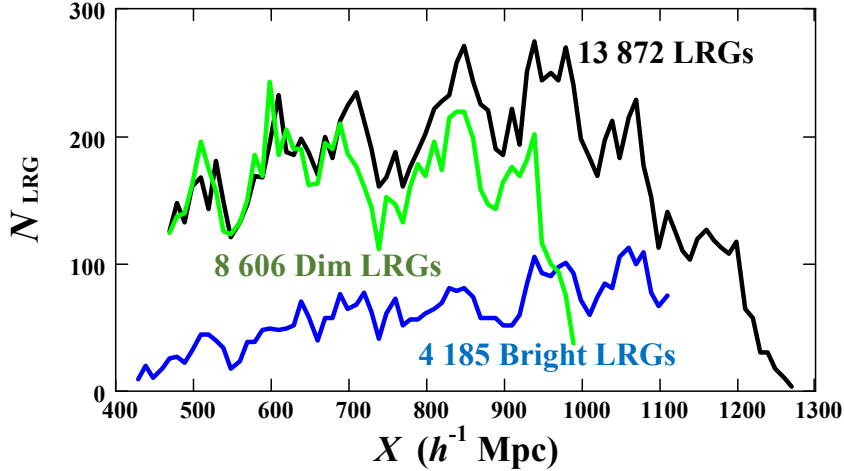}
\caption{
{\footnotesize
Three 
different   distribution functions 
$N_{\rm LRG} \equiv  N_X (X_0)$  
as a number of cumulative  projections  
on a selected  axis 
$X_0$ ($\alpha_0$, $\delta_0$;  see text)    
of the  Cartesian  LRG coordinates  
detected through six sectors (windows) 
in the sky shown in Figure~{\ref{BrPRk}}. 
Characteristics of all three
samples, DR7-Full  (black line), DR7-Dim (green line) and DR7-Bright (blue line),  
are given in Table~\protect{\ref{tab:samples}} }}   
\label{Distrib}
\end{center}
\end{figure}

\begin{figure}[t] 
\begin{center}
\includegraphics[width=0.47\textwidth]{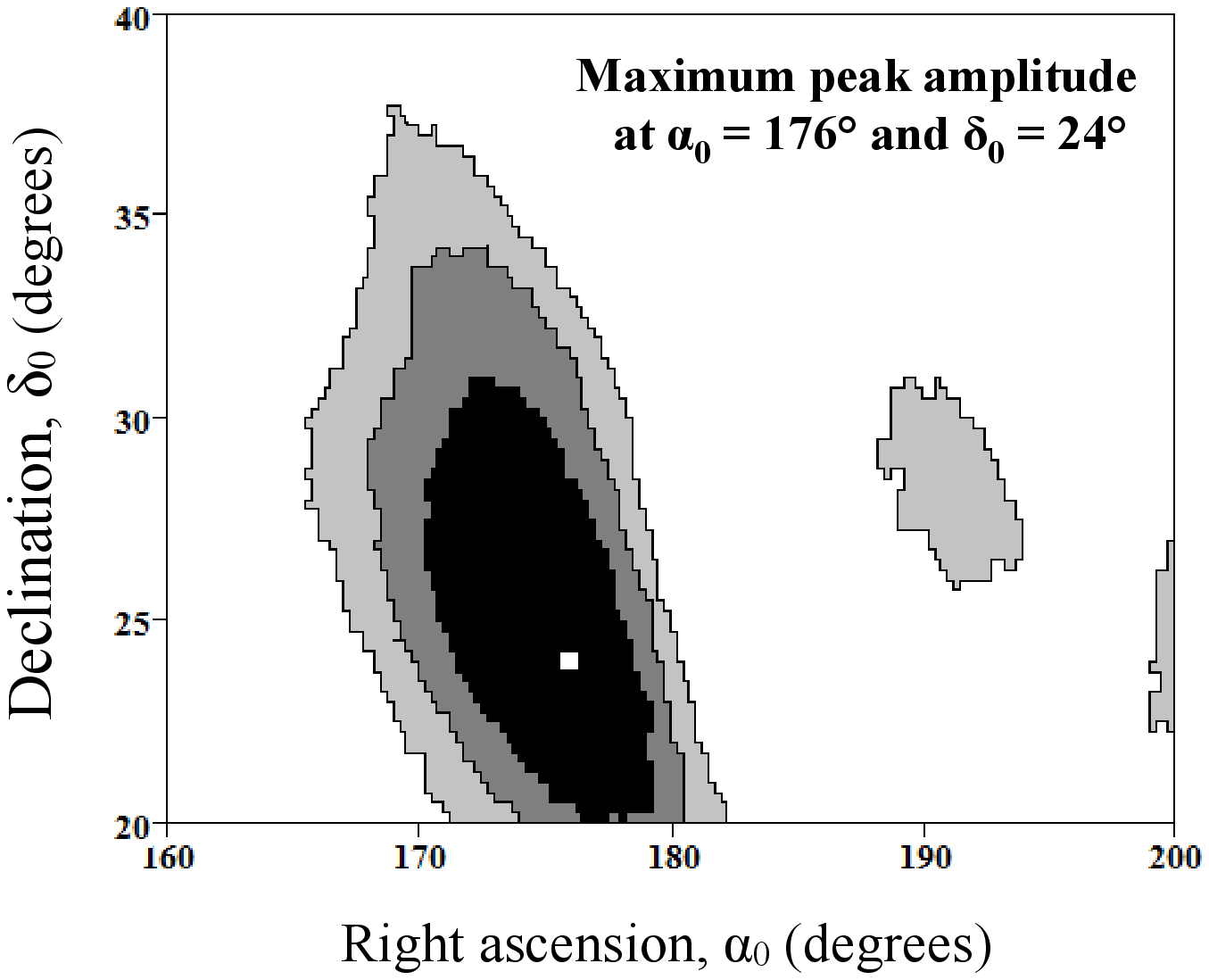}
\hspace{2mm}
\includegraphics[width=0.45\textwidth]{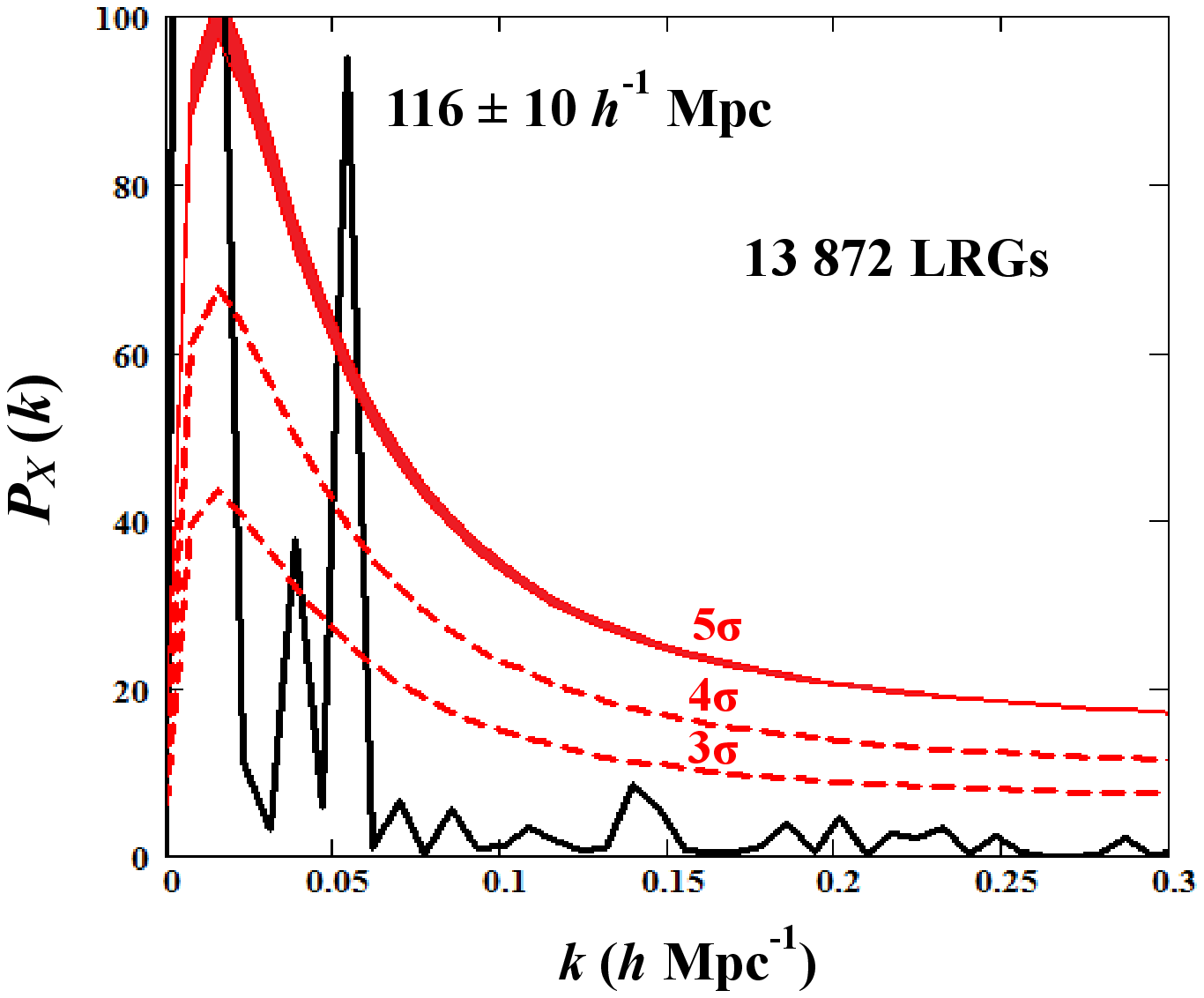}
\caption{
{\footnotesize
(\textbf{Left  panel}):
Areas of fixed confidence levels
of the peaks in 
1D power spectra $P_X(k)$  
at  $k_{\rm max} = 0.054~h$~Mpc$^{-1}$
calculated using Cartesian LRG coordinates  projected  
on various axes $X$; the data 
(corresponding to the sample DR7-Full)
are collected
for  the  six  sectors  shown
in  Figure~\ref{BrPRk};
points  in the $\alpha/\delta$-plane indicate 
the directions  of axes. 
The colored areas  are:
light gray {$-$}
$(3\sigma)$,
intermediate grey {$-$} $(4\sigma)$
and dark grey {$-$} $(5\sigma)$; 
the white  square indicates  the  direction of  the maximum
spectral peak height ($\alpha_0$ and $\delta_0$).  
(\textbf{Right  panel}):
1D power spectrum 
with maximum peak ($k=k_{\rm max}$) calculated  
for the projections on  $X_0$-axis
( $\alpha_0$,\  $\delta_0$) of
the  Cartesian  LRG coordinates  
registered  through  the  same   six  sectors; 
significance levels  $3\sigma$ and $4\sigma$
(dashed lines) and the  narrow band  $5\sigma$ 
are plotted similarly to the levels and band 
in the  right  panel of 
Figure~\protect{\ref{BrPRk}} (see text) } }
\label{PXk} 
\end{center}
\end{figure}

\begin{figure}[t]   
\begin{center}
\includegraphics[width=0.47\textwidth]{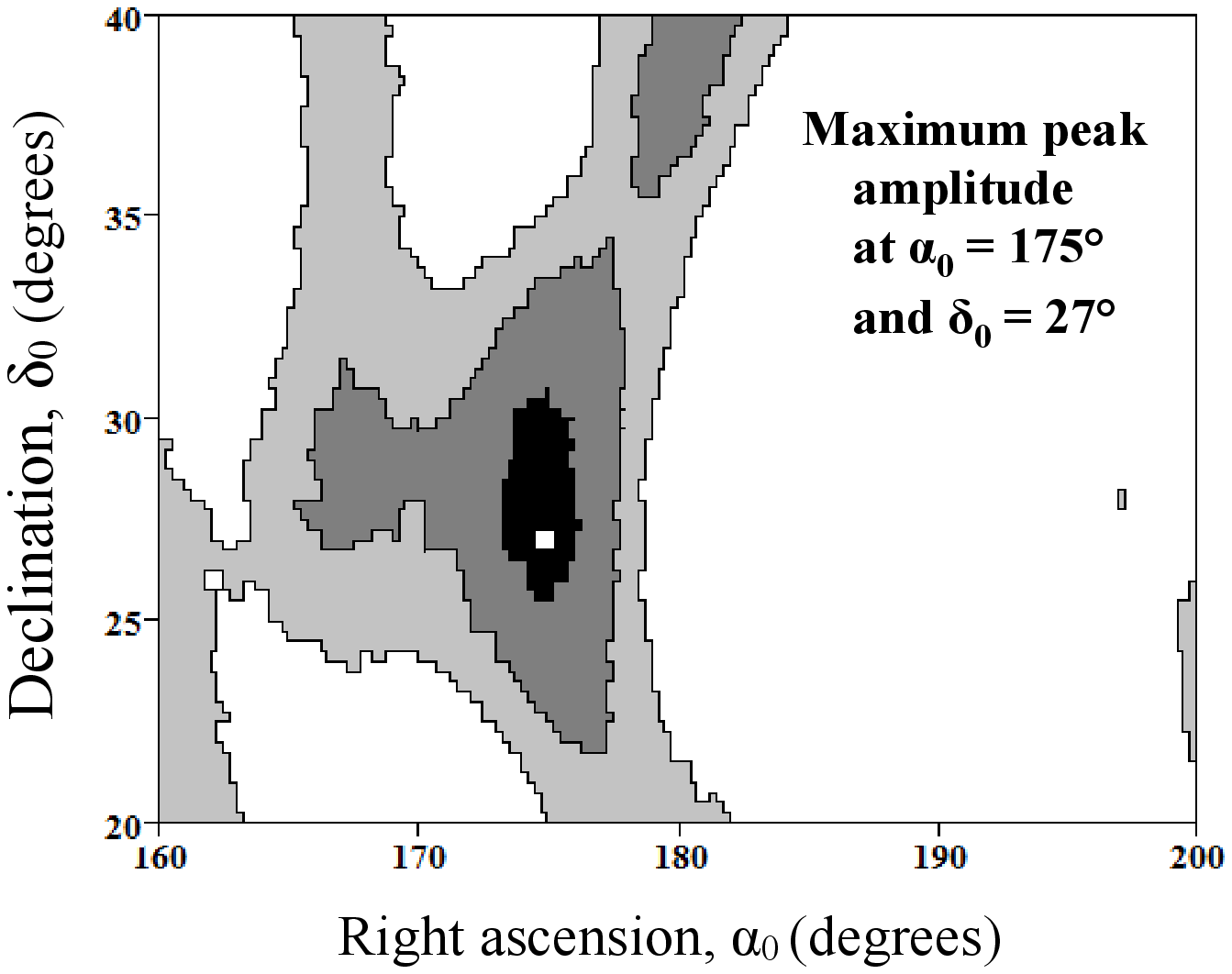}
\hspace{2mm}
\includegraphics[width=0.445\textwidth]{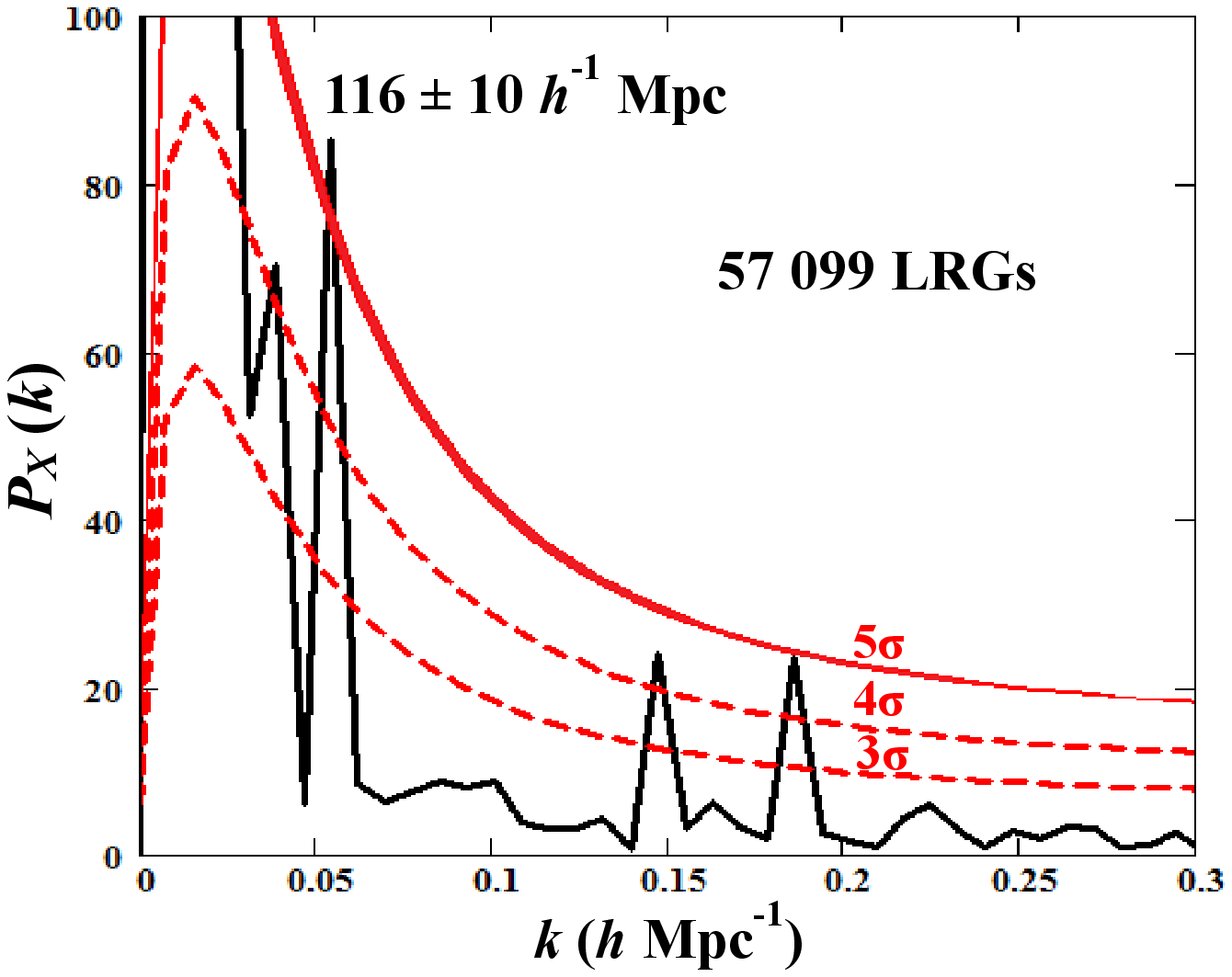}
\caption{
{\footnotesize
Same 
as the  left  and  right 
panels of Figure~\ref{PXk}, respectively,
but for all of the  data of the rectangle region shown  
in Figure~\ref{rectPRk}. The~maximum peak direction
($\alpha_0$ and $\delta_0$)
practically coincides (slightly differs) with
that in Figure~\ref{PXk} (see Table~\protect{\ref{tab:samples}}) }}
\label{rectPXk}
\end{center}
\end{figure}

\begin{figure}[t]    
\begin{center}
\includegraphics[width=0.44\textwidth]{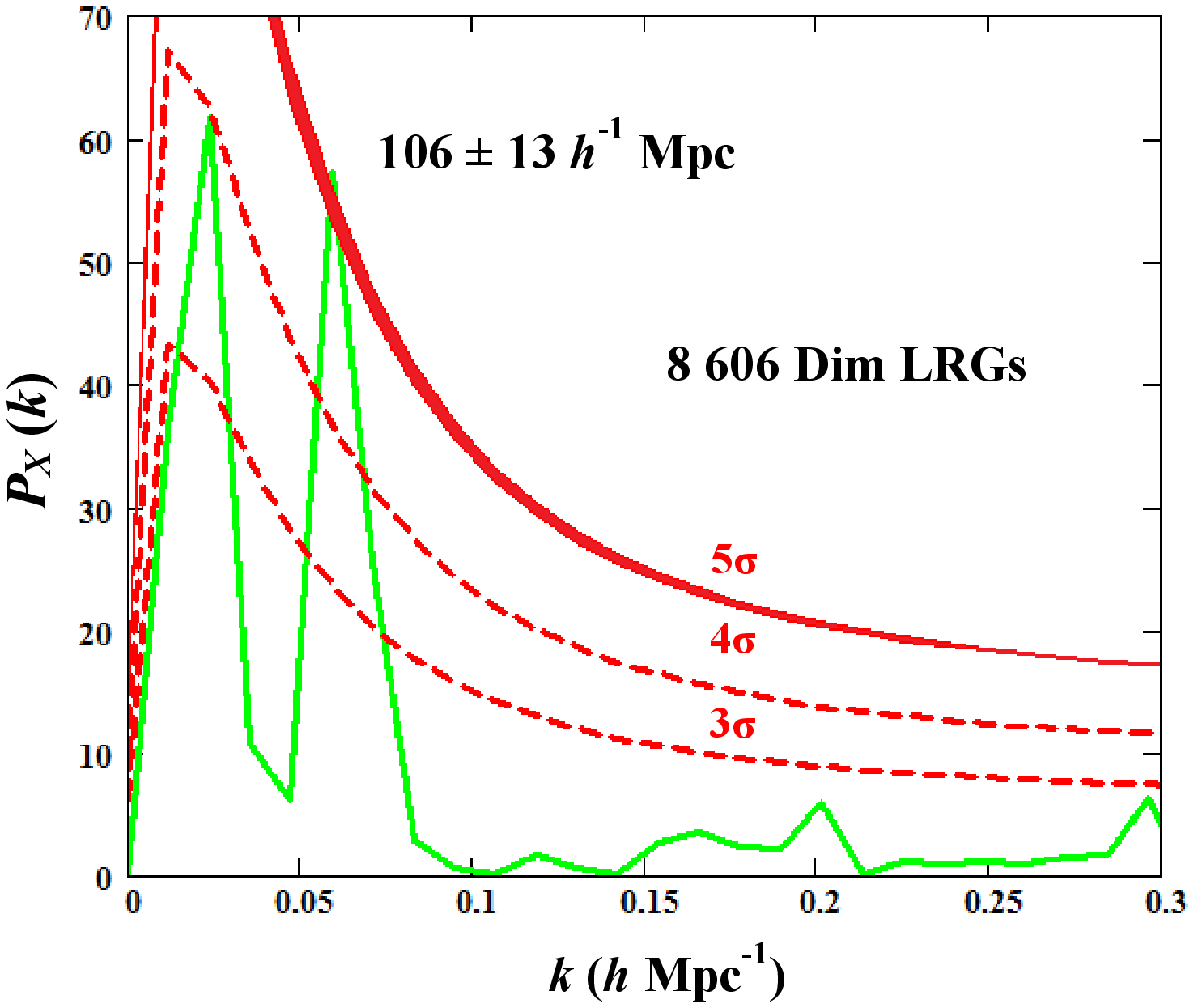}
\hspace{2mm}
\includegraphics[width=0.47\textwidth]{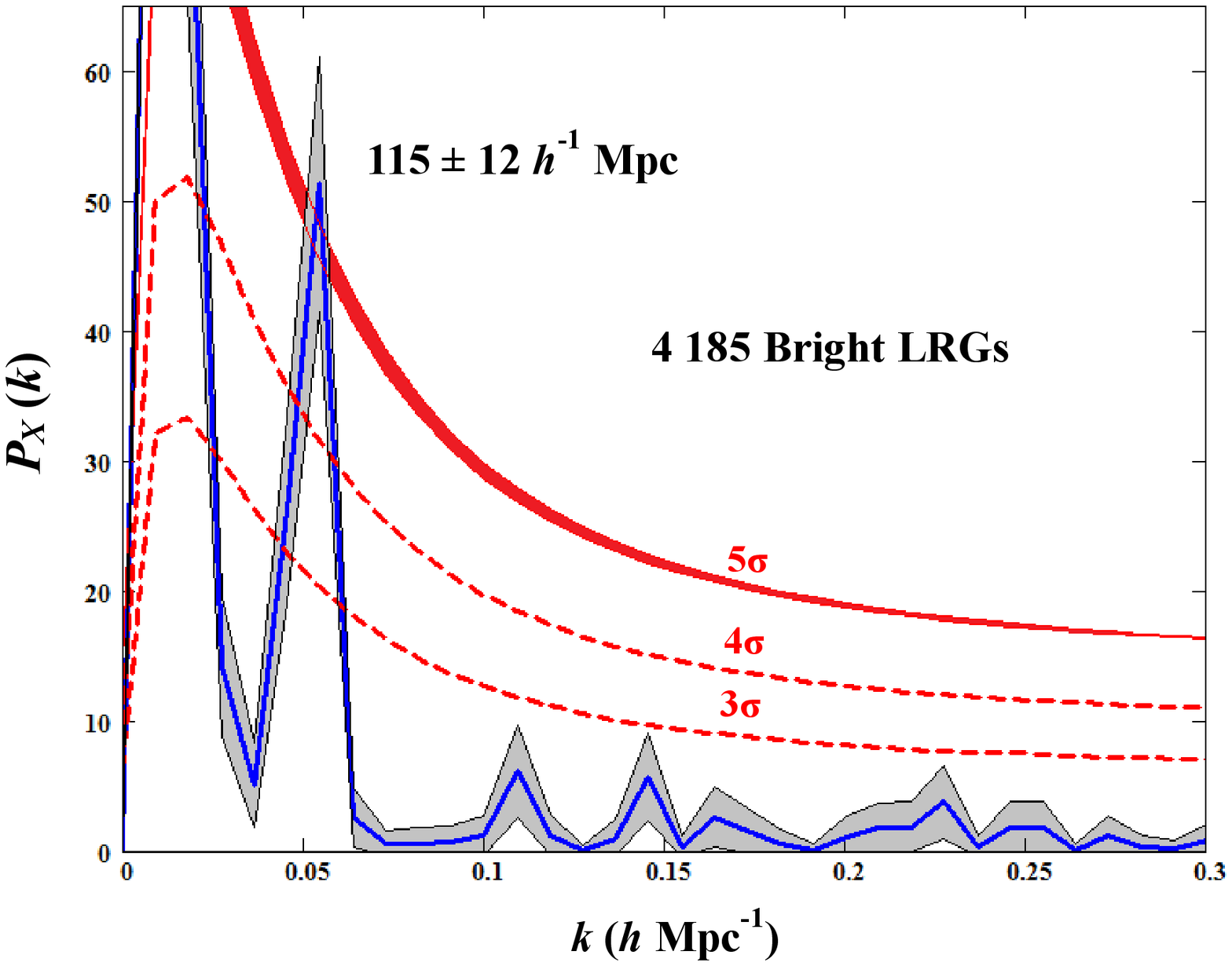}
\caption{
{\footnotesize
Same  
as the  right 
panels of Figure~\ref{PXk} 
but for the subsamples DR7-Dim 
(\textbf{left  panel}) and DR7-Bright 
(\textbf{right  panel}); 
see Table~\protect{\ref{tab:samples}} 
and text for details). 
The gray stripe  near the power spectrum,
plotted for DR7-Bright,  demonstrates error bars,
calculated by the jackknife technique }}
\label{Dim-BrightPXk} 
\end{center}
\end{figure}

Figure~\ref{Distrib} represents
three distribution functions $N_X (X_0)$
calculated as a number of cumulative projections 
on the axes  $X_0$ (see Table~\ref{tab:samples} ) 
of Cartesian LRG coordinates   
registered through  six sectors
shown in the left  panel of  
Figure~\ref{BrPRk}. 
It is seen that all three curves have a quasi-oscillating character, 
i.e., represent an alternation of peaks and dips,  the~positions 
of a number  of such features being  mutually consistent. 
%
The results of our calculations 
of the  1D power spectrum   
are represented in   Figures~\ref{PXk}--\ref{Dim-BrightPXk}. 
The first two of them are based on data of the sample DR7-Full, 
while the third one {---} on data of the subsamples DR7-Dim and DR7-Bright.
The first and third ones  relate to all six sectors discussed above.
The left  panel  in  Figure~\ref{PXk}  shows three confidence 
areas (shades of gray) on the sky indicating  peak amplitudes 
(for  the same  scale  $116~h^{-1}$~Mpc) exceeding
the significance levels $3\sigma$ (light gray), 
$4\sigma$ (darker gray)
and $5\sigma$ (dark), respectively.  The maximum value of the peak
is achieved  along  the direction of $X_0$ 
with coordinates $\alpha_0=176^\circ$
and $\delta_0=24^\circ$ (small white square).    
\unskip

The  right  panel represents the power spectrum $P_X(k)$ 
calculated for normalized 1D distribution  (\ref{NNX}) 
of  the LRG Cartesian  coordinate projections  
on  the $X_0$-axis with $\alpha_0$ and $\delta_0$ indicated above; 
the respective  sample size   is  13,872 LRGs. 
Two dashed lines  show
significance levels  $3 \sigma$ and $ 4\sigma$,
while the narrow band 
(taking into account error bar $\pm 1 \sigma$) 
corresponds to  the level $5 \sigma$
(in the case of  Figure~\ref{PXk}, we  get  $A=156 \pm 6$), 
which are calculated in the same manner 
as described in Section~\ref{sec2}
with the use of  LasDamas catalog (``lrgFull-real'') 
for the same six  sectors in the sky.
We compute  the   ${\cal N}_{\rm LD} = 80$ 
power spectra $P_X^{\rm LD} (k)$ 
produced for the normalized 
distributions   (\ref{NNX}) along  the  axis  $X_0$.
As a result, one can see
that the dominant peak of our special interest 
far exceeds the level of $5 \sigma$  
and demonstrates an amplitude of
about 100. This amplitude is noticeably larger than  
the similar  peak in  Figure~\ref{BrPRk}. 

Fourier analysis of these 
quasi-periodical components 
carried out separately for each selected  sector 
shows that phases of the two
most significant  sectors
(No 3 and 5 in the left  panel of Figure~\ref{BrPRk}, 
see also Table~\ref{tab:sectors})  
get  closer relative to 
the case of radial distributions.
This phase convergence provides  a major  
contribution to the cumulative spectrum of 
the  six sectors  in~total.

\unskip
Figure~\ref{rectPXk} is organized similarly to Figure~\ref{PXk}
but represents  the results of calculations  of 1D power spectra
produced for the whole rectangle region shown in the  left 
panel of \mbox{Figure~\ref{rectPRk}}. 
The direction of the maximum amplitude
of peaks in the power spectra $P_X (k)$ is only slightly shifted
relative to the case of the  six  sectors in Figure~\ref{PXk}, 
i.e., $\alpha_0=175^\circ$  and $\delta_0 = 27^\circ$. 

The  right  panel  represents
the  1D power spectrum calculated along $X_0$  for  a 
sample of 57,099 LRGs.  One can see a strong peak at the same 
period  $116 \pm 10~h^{-1}$~Mpc  
but with an  amplitude  a bit  lower than 
in the previous case. The~significance levels  
(dashed lines  and a narrow band) 
are constructed
similar  to the  right   panel in Figure~\ref{PXk} using LasDamas 
catalog  (``lrgFull-real'') but  for the whole rectangle region in the sky.
This means that the proposed periodical structure oriented
along   $X_0$  can manifest itself
even for the entire rectangle  area under 
consideration (cf. with the  right  panel of Figure~\ref{rectPRk}).
 
In both  Figures~\ref{PXk} and \ref{rectPXk},
we can notice a smaller but also significant peak at lower
$k < 0.05~h$~Mpc$^{-1}$  and two significant peaks 
at $k  \la (0.15$--$0.2)~h$~Mpc$^{-1}$   in Figure~\ref{rectPXk}. 
These features  may indicate a more complex character of the 
structure under discussion  than  a single periodical dependence on $X$
but with a dominant role of  the one  \mbox{highlighted  component}. 

\unskip

Figure~\ref{Dim-BrightPXk} is 
organized in the same way as  in the right panels in Figure~\ref{PXk} 
and   deals  only   with  the  data  of  
six chosen sectors. 
The figure is  plotted  for relatively more 
homogeneous  samples DR7-Dim and  DR7-Bright;  the latter 
contains relatively fewer statistics.
With this in mind, 
we slightly extended the low boundary of the $X_0$-projections
in the case of DR7-Bright  (at fixed $z=0.16$ 
as  noted in  the  footnote  No. 10), 
to increase the amplitude of  the peak. 
However, we can argue that such  variations of the boundary  
do not diminish  the significance of the peak below $4\sigma$.

Moreover, following 
[50],
we applied the jackknife procedure
for the calculation of
power spectrum error bands,
obtained
for DR7-Bright data (right panel).
The stripe takes into account random variations
of the data  used for the calculation of the 
power spectrum. It can be seen that the errors 
cannot drastically affect the main peak
significance. 

Let us note that  
when considering  the  sample  DR7-Bright,
we do not use the reduction 
procedure (\ref{reduc}) for  calculating significance levels
in the  right   panel of  
Figure~\ref{Dim-BrightPXk},
because  the trend  DR7-Bright  
turns out to be  quite  similar  to  the  trend  of  mock
LD data (catalog  ``lrg21p8-real'')  selected  
under  the  same  spatial  conditions. 
This confirms  the  assumption 
that the reduction procedure 
introduced  in Section~\ref{sec2}
does not  significantly  affect the  position
 and magnitude  of the main  peaks 
in the power~spectra. 

It is also worth noting that, based on the DR7-Bright sample, 
we compare  two $N_X(X_0)$  distributions constructed for the  six sectors, 
as indicated in   Figure~\ref{Distrib}, 
and for the entire rectangular region shown in Figure~\ref{rectPRk}.  
Both distributions are similar and represent an alternation 
of peaks and dips; 
however, the~amplitudes of these peaks and dips for the six sectors 
turned out to be  a bit larger, which indicates some advantage 
of these sectors in tracing an assumed  structure.
The correlation coefficient of two curves is   0.51, 
which exceeds level  $4 \sigma$ 
for the considered volume of~samples.

We can summarize
that the celestial coordinates of  the axes $X_0$  
are quite close  in all four  cases  under consideration 
(see  Table~\ref{tab:samples}). 
Moreover,  the~position  and significance of the peaks 
in the  right  panels of  Figures~\ref{PXk} and \ref{rectPXk}  and 
in both panels of Figure~\ref{Dim-BrightPXk} are also mutually  consistent.
Thus,  noticeable changes  in the statistics  and  homogeneity of the samples
do  not   significantly  change  the results, 
confirming  their~robustness.

\begin{figure}[t]    
\begin{center}
\includegraphics[width=0.47\textwidth]{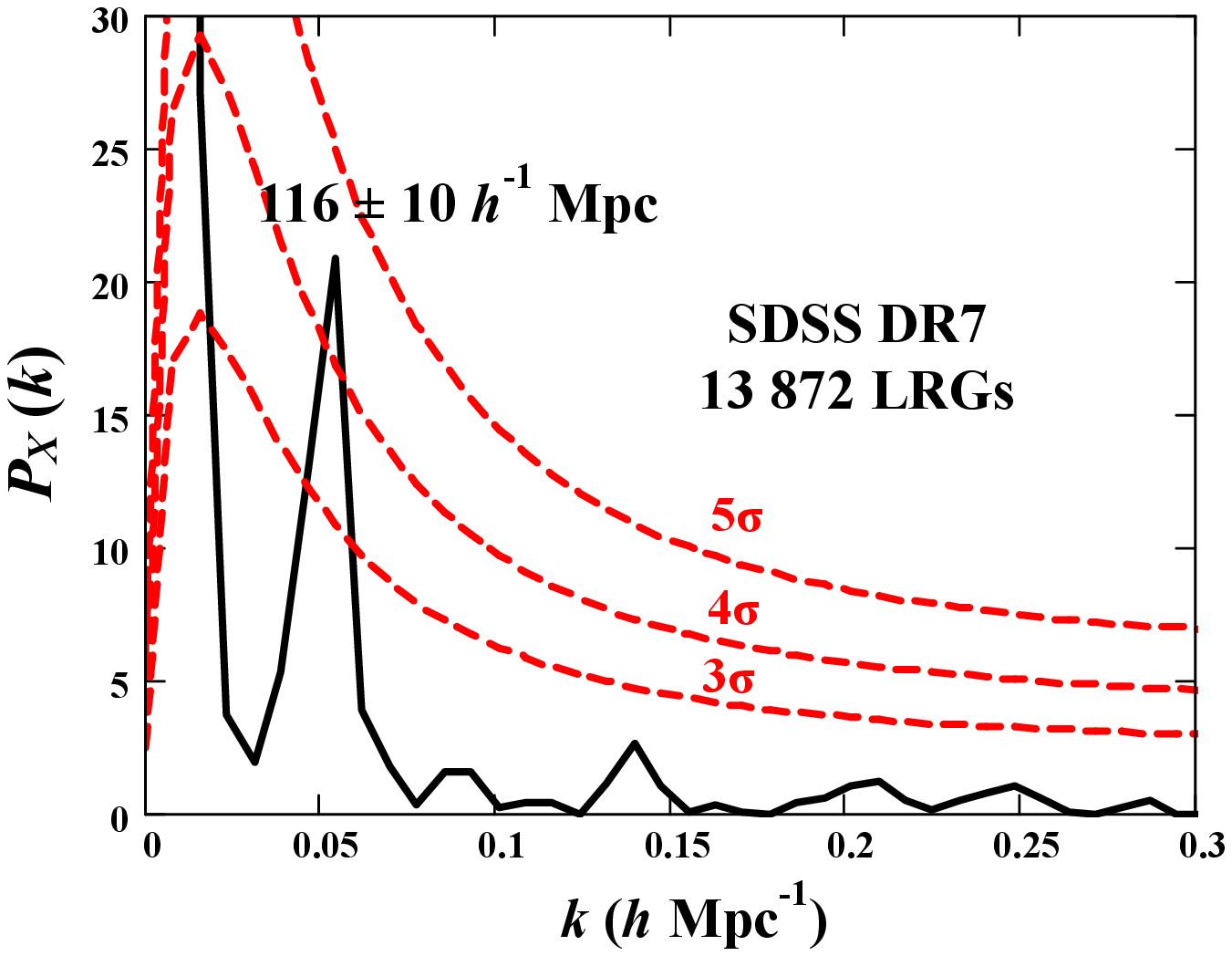}
\hspace{2mm}
\includegraphics[width=0.42\textwidth]{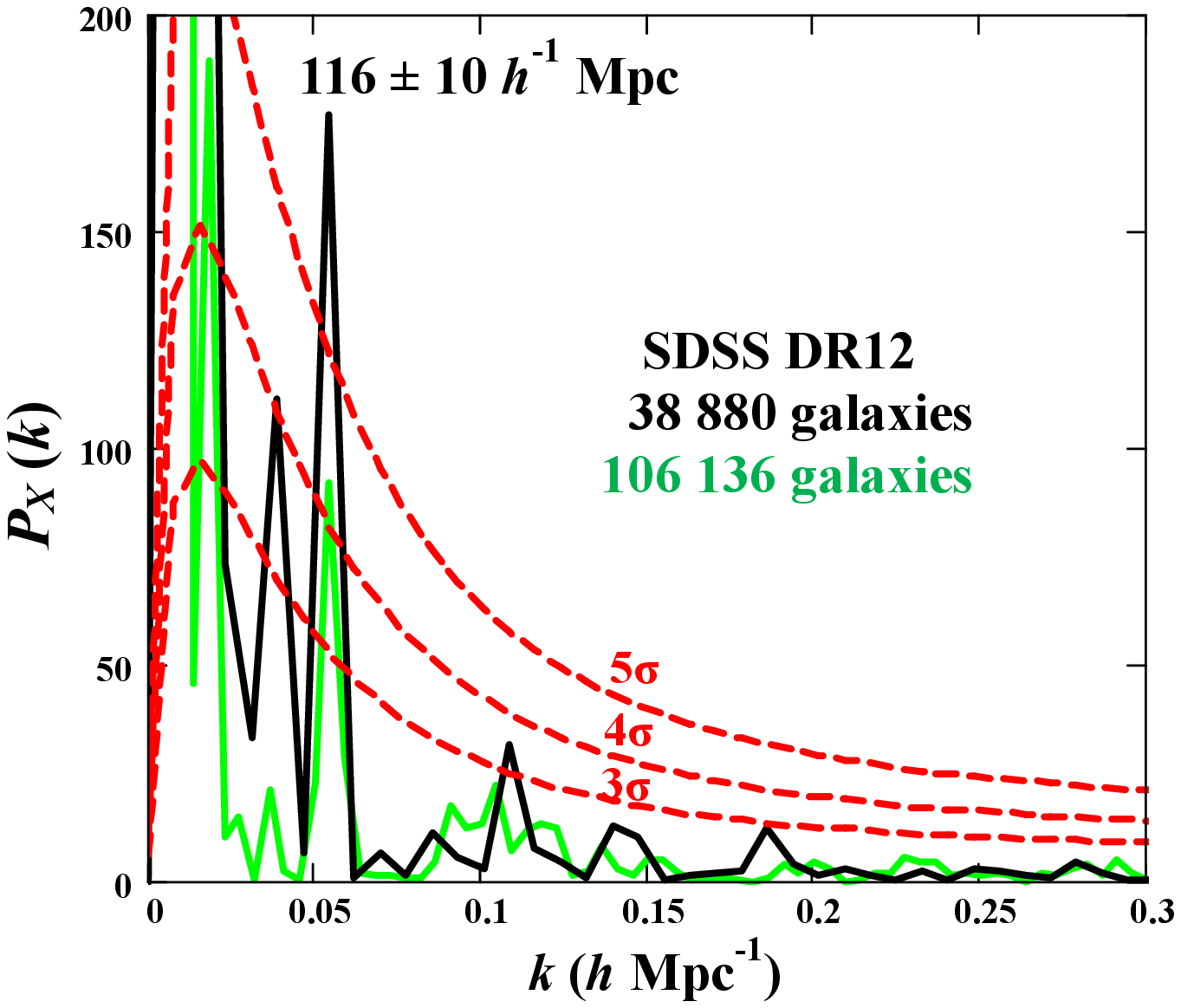}
\caption{
{\footnotesize
(\textbf{Left  panel}): 
Same as the  right 
panels in  Figure~\ref{PXk} 
but with the inclusion of the Hann  function 
\protect{(\ref{Hann})} in the calculations 
of the window Fourier spectra  
(using  the same DR7-Full data). 
(\textbf{Right  panel}):
Same  as  in the  right  panel  of  Figure~\ref{PXk}  
but  using   the data of SDSS DR12  
related  to  the same six sectors in the sky. 
The green  curve  is the  1D  power spectrum 
calculated for the  extended  interval   
$464 \leq  X  \leq  1844~h^{-1}$~Mpc 
$(0.16 \leq z \leq 0.72)$
of projections of the Cartesian galaxy coordinates 
on  the  axis  $X_0$
($\alpha_0=176^\circ$ and $\delta_0=27^\circ$,\
see Table~\protect{\ref{tab:samples}}) 
corresponding to  the  maximum peak 
at   $k_{\rm max}  = 0.054~h$~Mpc$^{-1}$.
The  black curve
is also a 1D power spectrum with the same $k_{\rm max}$  but 
calculated for the  smaller 
interval  $464 \leq  X  \leq  1274~h^{-1}$~Mpc
$(0.16 \leq z \leq 0.47))$
used in Figures~\protect{\ref{PXk}} and  \protect{\ref{rectPXk}}; 
in this case, $X_0$  is oriented along direction   
$\alpha_0=175^\circ$ and $\delta_0=25^\circ$;
the sample sizes for both  sets of data are also shown.
The significance levels $3\sigma$, $4\sigma$  and $5\sigma$
(dashed lines ) 
are calculated for  the extended sample of SDSS DR12 
(see Section~\ref{sec5}) } }  
\label{wf_DR12} 
\end{center}
\end{figure}

In support  of  this statement, we introduce 
two auxiliary  panels
in Figure~\ref{wf_DR12}  showing weak dependence  of  the  
results  on the degree of  sample homogeneity.   
Indeed,  the~left panel of Figure~\ref{wf_DR12} 
shows the effect of the window  Fourier transform on the
power spectrum obtained for  the same distribution
of the  Cartesian  coordinate projections 
on the axis $X_0$   as  in  the right panel of Figure~\ref{PXk}. 
As a window function, we use the Hann function  (e.g.,  [50, 51]): 

%
\begin{equation}
{\cal W}_l = {1 \over 2} \left[ 1 - \cos \left( {2 \pi X_c^l \over L_X} \right) \right],
\label{Hann}
\end{equation}
notations  on the right-hand side 
are the same as in Equation~(\ref{NNX}). 

The function
(\ref{Hann}) smooths the distribution of objects (points) 
along the edges
of  considered intervals, 
thereby suppressing the influence of sample inhomogeneities
(visible, e.g.,~in Figure~\ref{Distrib}  for DR7-Full data)
and smoothing out  spurious periodicities 
induced by the boundaries of  the  intervals  (e.g.,  [50]). 
On the other hand, this function strongly 
suppresses some of the useful information  and,  in particular, 
reduces traces of the periodic structure,  if~it is present, 
in the power spectrum (e.g.,  [51]). 

{We multiply} 
the function (\ref{Hann}) 
by the normalized distribution  Equation~(\ref{NNX}), 
perform the Fourier transform of this product and
construct the power spectrum following 
the modification of  Equation~(\ref{PRkm}) 
for projections onto the X-axis, as~described  above.  
Similarly, to~obtain the significance levels shown 
in the left panel of  Figure~\ref{wf_DR12},
we perform the same  window Fourier transform  
of the one-dimensional distribution 
derived from  the  LasDamas  data   
(the same  catalog ``lrgFull-real''). 
\unskip
Comparing  the 
left panel in Figure~\ref{wf_DR12} 
with   the right panel in Figure~\ref{PXk},
one can notice  the significant single peak at the same 
$k_{\rm max}=0.054~h$~Mpc$^{-1}$
but  its  amplitude is greatly reduced due to the influence 
of the window function on the power spectrum.  
However, the~significance levels  are also greatly reduced 
(slightly less than the peak amplitude),
and the resulting significance does not fall below $4 \sigma$,
i.e., the  peak remains quite significant.

\section{Traces of Spatial Structure in SDSS~DR12}\label{sec5}
To  verify the results of Section~\ref{sec4},
we  probe  the existence of a  quasi-periodical structure on the basis of
significantly expanded statistics of  cosmologically  distant   galaxies  accumulated  
in SDSS  DR12. We consider these calculations as  pure  preliminary  ones
because here  we   do not 
take into account 
nonhomogeneity  and selection 
effects  of used  data, 
do not study  
the wider  sky area  available for  DR12,  and
in particular,   are not looking for 
additional  special   
sectors in the expanded  region
with significant  spectral features, 
etc. 
Such calculations  are  the subject  of  future work.
Here our task is only to establish whether 
there  are contradictions between power spectra obtained 
along certain  directions  for  quite different  statistics
presented in  DR7 and~DR12. 

We employ  the  data of  DR12 accumulated 
only for the northern  hemisphere  
in the sky  and  collected  in two  files: \\
(1)\   {\footnotesize{\it galaxy\_ DR12v5 \_ LOWZ \_ North.fits.gz}},  \quad  
(2)\  {\footnotesize{\it galaxy \_ DR12v5 \_ SMASSLOWZE3 \_ North.fits.gz}} \\
which are
available in the Science Archive Server 
\footnote{\it {https://data.sdss3.org/sas/dr12/boss/lss/} }.
A description of the catalogs  DR12 can be found, e.g., in
[52-54].

One can  produce  calculations similar 
to those that  are  performed  to plot 
Figure~\ref{PXk}  
using data related to the same six sectors
in the sky, as  presented 
in Table~\ref{tab:sectors}.
We consider  both  an  extended interval of redshifts   
$0.16 \leq z \leq 0.72$ ({\it   SMASSLOWZE3})  and
the same  (shorter)  interval  $0.16 \leq z \leq 0.47$ ({\it   LOWZ})
as it  is  studied in  Sections~\ref{sec3} and  \ref{sec4}.  
However, now our total sample contains 106,136  galaxies  
for the extended interval and 38,880 galaxies in the 
shorter  interval (instead of 13,872 LRGs  in Figure~\ref{PXk}). 

At first, we consider the extended redshift interval. 
As in the previous section, we choose the initial Cartesian CS
and produce a lot of  CS  rotations
within the same area in the sky $160^\circ \leq \alpha \leq 200^\circ$ and 
$20^\circ \leq \delta \leq 40^\circ$, as in Section~\ref{sec4}. 
In this case, we restrict ourselves 
to a fixed  interval  $464 \leq  X  \leq 1844~h^{-1}$~Mpc 
along each direction,  
similarly  to how it is  done  for   the  DR7 data.
For each direction of  the $X'$-axis,
we also  calculate the discrete analog of  Radon transform, 
i.e., we summarize all the galaxies  from a volumetric sample 
whose projections of the initial (fixed) Cartesian coordinates fall into a given
bin along the rotating  axis. 
In such a way, we obtain a  1D distribution  $N_X (X_c)$, where $X_c$ 
is  the  central  point  of  a   bin and,~using Equation~(\ref{NNX}), produce 
the normalized  distribution  ${\rm NN}_X (X_c)$.

This allows us to calculate the power spectra $P_X (k)$
for  various  $X'$-axes and to find the  direction
of the maximum peak  amplitude  at  $k_{\rm max}=0.054~h$~Mpc$^{-1}$. 
An example  of  such calculations 
for the direction of the maximum  amplitude 
($\alpha_0=176^\circ$ and $\delta_0=27^\circ$) 
is shown  in the right panel of Figure~\ref{wf_DR12}  
by the  green  curve.
One can see that  these coordinates   are fairly  close   to
the  similar $\alpha_0$ and  $\delta_0$ in  Figure~\ref{PXk} 
(see also Table~\ref{tab:samples}).  
The  curve  manifests a  moderately  high amplitude 
of  the  peak   with a significance    of $\sim$$4 \sigma$.

Three  dashed lines  in the right panel of  Figure~\ref{wf_DR12}
show significance levels $3\sigma$, $4\sigma$
and $5\sigma$, respectively. The~calculations of these levels
are  carried out  on the basis of the total sample   (106,136 galaxies)
by constructing a large number (${\cal{N}}_X$ = 861) 
of power spectra for various 
directions of $X'$  evenly covering the area in the sky
under investigation.  Following  \mbox{Section~\ref{sec2}}, 
we obtain an averaged power spectra  $\langle P_X (k) \rangle$, 
and using the Equations~(\ref{Cij})--(\ref{fCDM}), \mbox{(\ref{L})
and  (\ref{calF})},  where all  indices ${\rm LD}$  are  replaced by $X$,
 we calculate the significance levels plotted in Figure~\ref{wf_DR12}
(right panel).

For comparison, the   black  curve demonstrates 
another  example  of  the
power spectrum calculations in the same 
interval   $464 \leq X \leq 1274~h^{-1}$~Mpc  as
it  was used in Section~\ref{sec4}. In~this case,
the full number of galaxies related to all six sectors 
reaches 38,880.   As~a result, we obtain
a slightly different direction of the maximum  amplitude 
($\alpha_0=175^\circ$ and $\delta_0=25^\circ$) 
at  the same $k_{\rm max}=0.054$. 
One can see that the amplitude of the dominant peak 
at the same $k_{\rm max}$  increases by approximately two times.
On the other hand,  the~decrease of the peak amplitude 
for  the  extended  interval of $X$ 
could be  a consequence of the limited  size of 
the quasi-periodical structure (if it exists) along  the $X$-axis.

Thus,  our  preliminary analysis  of  DR12 data confirms an  appearance
of the significant   feature at  $0.05 <  k  <  0.07$ in the power spectra
and thereby the  possibility  of  existence of 
the  quasi-periodic component oriented
along the highlighted directions. Let us note
that  the narrow bunch of  directions $X_0$  found  in our study  
passes through the origin of CS (observer)
by construction. However,  we suppose that  the real 
axis of periodicity could be  (quasi-)
parallel  to the found axis (bunch of axes)   
and  probably  be shifted in space, 
so that  the  Radon transform 
does  not change (see Section~\ref{sec4}).

\section{Conclusions and~Discussion}\label{sec6}

The focus  of this work is  the search for 
traces of  the anisotropic  quasi-periodic structure
in the spatial distribution of cosmological distant  galaxies  and 
application of a  proper  method  for assessment   of  its   significance.  
Summarizing the results obtained in \mbox{Sections~\ref{sec3}--\ref{sec5}}, we can    
hypothesize that at the  considered  redshifts, mainly at
$0.16  \leq z  \leq 0.47$
(see also Table~\ref{tab:samples} for details),
a large elongated quasi-periodic
structure could exist 
with characteristic  scale  
$\Delta X = 116 \pm 10~h^{-1}$~Mpc.

In order to specify  the 
main axis  of the structure,
we perform the discrete 3D Radon transformations
along  various  axes $X$ and calculate 
the power spectra   for  corresponding  1D distributions. 
One can imagine  that  the real axis of 
quasi-periodicity  (if it exists)
can  be  parallel to this  axis  
but  arbitrarily  shifted in space.
Such an approach may be treated as an application 
of the computer tomography elements  to the analysis of
the large-scale inhomogeneities of the matter 
(e.g., [55])
and  possible  signs  of  their~quasi-periodicity.

Along   special  directions  located 
within  the  narrow  intervals of 
equatorial coordinates 
$\alpha_0 \simeq  175^\circ  - 177^\circ$  and 
$\delta_0 \simeq  22^\circ - 27^\circ$,
the structure is likely to have a  maximum  
scale of  $\ga$800~$h^{-1}$~Mpc.
Our estimations 
show that a signal-to-noise ratio 
for  the dominant   spatial oscillations
averaged over all  selected  directions  indicated  
in Table~\ref{tab:samples}
turned out to be   $\sim$$2.0$,
while  a density contrast is $\sim$$0.1$ \footnote{    
For these estimates, we use a modification of  Equations~(\ref{fCDM}) and (\ref{Tq}) 
of  [42]
which we especially tested by simulations, 
namely, the~averaged density contrast  
$\langle \delta \rangle$$\sim$$\langle \sqrt{ 4 P(k_{\rm max})/ N_{LRG} } \rangle$ 
and signal-to-noise ratio  
$\langle S/N \rangle$$\sim$$\langle \sqrt{4 P(k_{\rm max})/ {\cal N}_b} \rangle$,   
where $P(k_{\rm max})$ is an amplitude  of the main peaks in the power spectra 
at $k=k_{\rm max}$,
$N_{LRG}$ is a volume of  samples, and
${\cal N}_b$ is a number of bins accepted for each direction $X$.}.

Among currently known large-scale structures 
in the spatial distribution of matter,
the  structure proposed here 
can be compared  to
the so-called Great Walls. To~our knowledge, up to now  
a few   Great Walls  have been reliably  established,
and their number is constantly growing.
These are the  CfA Great Wall 
[56],
Sloan Great Wall  
[57, 58],
BOSS Great Wall~[59]
and Saraswati wall-like structure  
[60].
One can also  refer  to such objects as the supergalactic plane 
[61]  
in the local Universe,   
as well as  the 
Sharpley Supercluster 
(e.g.,~[62, 63]),   
and the recently revealed  South Pole Wall  
[64].

The structure proposed in the present
work  is about  two 
or more  times larger along  the major  axis  
than the appropriate  scales 
of the Great Walls.   
Relative to  $z$, 
the proposed structure is  situated somewhere
between the BOSS ($0.43 < z < 0.71$) 
and the  Sloan Great Walls ($0.04 < z  < 0.12$), 
and   includes    redshifts  ($z \approx 0.3$)   
of the Saraswati Wall.   
The major  axis  of  the assumed  structure  
is  directed  relatively close to  an area in the sky
 ($152^\circ \la \alpha \la  170^\circ,\
44^\circ \la  \delta  \la  58^\circ $)   
where  the Boss Great Wall  is  located.
The peculiarity  of the  structure  discussed here  is in its 
quasi-periodical character  with low  amplitude  (or overdensity), 
which could  be revealed only 
by specific techniques such as the anisotropic Fourier analysis
employed in the present~work.

In this regard, the~question arises 
about the largest  allowable  scales
of cosmological structures
(huge superclusters and voids between them) consistent with the generally 
accepted $\Lambda$CDM cosmological model 
(e.g.,  [60]) 
with its extensive observation base. 
A possible answer to this question
was given  
in  [65],
where  it  was   
shown  that the largest 
structures of relatively low density can reach
several   hundred  Megaparsecs without conflicting with 
the isotropy and uniformity of the 
$\Lambda$CDM  model. 
However, in~the literature,   there is evidence 
(e.g.,  [66])
that the size of inhomogeneities 
in the distribution of cosmological matter can be  much larger. 
Our estimates also show significantly larger dimensions 
of the anisotropic  quasi-periodic structure. 
It means that 
the compatibility issue
of such a scale with the currently dominant $\Lambda$CDM model
remains~open. 

The same applies to the question  of origin of
the structure  that  we assume. It is quite likely, as~suggested 
in  [30],
that the origin of anisotropic
quasi-regular structures is associated with phase transitions
during the inflation epoch or immediately after it.  A possible
way of solving the problem of the emergence 
of quasi-periodic structures in the early Universe 
was indicated in a recent work  
[67].   

In addition, let us note  that  sector No~5 
represented  in the left  panel of Figure~\ref{BrPRk} 
(see  also  Table~\ref{tab:sectors})
contains the  direction to the north Galactic pole. 
In principle, the~structure assumed here  
might  be  consistent with
the  pencil-beam quasi-periodicity 
at a scale $\sim$130~$h^{-1}$~Mpc
found in the pencil-beam surveys 
near  both Galactic poles  
(see   [17] and  
a few  references  in the Introduction).  
It should be mentioned, 
however,
that these  results  were   
criticized in the literature
(e.g., in  
[68-70])
and perhaps the significance  of periodicity 
was overestimated  by the authors,   
who  used 
the statistics available to~them.

Actually,
our analysis is 
different from the pencil-beam  treatment,
although in principle it may not contradict  it.   
We analyze the projection of a volumetric  array of points
(LRGs) on  selected directions,  whereas the pencil-beam analysis
has been  produced  for the 
distribution of galaxies along a set of 
narrow observational  cones.
Note that the directions along which
there are significant peaks in the power spectra
do not coincide with the direction to the Galactic pole.
Moreover, Figure~\ref{rectPXk} shows 
the significant peak in
the power spectrum calculated 
for a large number of LRG projections 
that were observed over the entire rectangular 
area in the sky, i.e.,~for LRGs   
collected from a huge spatial~volume.

The structure proposed in the present work 
has a ragged character 
and can only appear in certain directions;
moreover, different  directions may  trace over
different  visual space periodicities, as was 
found, e.g.,~in
{[29] and  [30]}.
Nevertheless,
two types of  periodicities 
(one based on the pencil-beam analysis  and
the other one discussed in this article) 
might  be interconnected,
such  as  two different probes of  the same complex 
quasi-regular~structure. 

Another point worth mentioning  here is
the closeness of the scale obtained  in this work  to
both  the  characteristic scales of 
the quasi-regular 
structure  formations and
the BAO phenomenon 
(see  Introduction).   
In our case,
we are most likely  dealing with an oriented anisotropic 
structure similar to those obtained  in
[17, 29] and  [30].
Moreover, it seems  to be   plausible
that the quasi-regular structure  could 
manifest itself  as several observed oscillations 
in space with a fixed  scale   (e.g.,  
[17, 23, 30]),  
exactly as   we   find  in this~work.

As for the BAO, one can expect
(see, e.g.,~[6])
that the primary perturbations in real space,
associated with oscillations in k-space, 
could  spread  relative to  their  original  
centers   (scattered  isotropic)
only within one acoustic wavelength. 
Such a concept of BAO  has been
confirmed to a certain degree 
by the calculations of  
[33].
Nevertheless,
it is also possible  that the proximity of the scales 
(for  all their differences) of  both  types of 
phenomena under discussion  is not accidental,  and~
they  have common progenitors
in primary perturbations at  the early stages 
of the Universe~evolution. 

In  any case,   our hypothesis  requires further detailed statistical studies,
including  an analysis of systematic errors that can lead to distortions 
of power spectra. This is especially true for DR12 data (Section~\ref{sec5}), 
which must be used with great care. 
Our approach could be
justified to some extent  by the fact that the use of several samples
with varying degrees of data heterogeneity
and with varying degrees of accounting for systematic effects,
leads to stable results.
Nevertheless, it should be remembered that
even with good statistics,
low-quality data
can lead to unreliable conclusions (see, e.g.,~[71]). 
In any case,  all the effects  considered  here require further
research, including the use of other catalogs 
of observational data. 
Moreover, these studies should be extended also to other  areas 
in the sky,  including the region of the south Galactic~pole.

\section{Acknowledgments}
The authors are deeply grateful to the anonymous reviewers for many helpful 
and constructive~comments.

{}
\end{document}